\documentclass[preprint,showpacs,preprintnumbers,amsmath,amssymb,prb]{revtex4}

\usepackage{graphicx}
\usepackage{dcolumn}
\usepackage{bm}
\usepackage{epsf}

\newdimen\tableauside\tableauside=1.0ex
\newdimen\tableaurule\tableaurule=0.4pt
\newdimen\tableaustep
\def\phantomhrule#1{\hbox{\vbox to0pt{\hrule height\tableaurule width#1\vss}}}
\def\phantomvrule#1{\vbox{\hbox to0pt{\vrule width\tableaurule height#1\hss}}}
\def\sqr{\vbox{%
  \phantomhrule\tableaustep
  \hbox{\phantomvrule\tableaustep\kern\tableaustep\phantomvrule\tableaustep}%
  \hbox{\vbox{\phantomhrule\tableauside}\kern-\tableaurule}}}
\def\squares#1{\hbox{\count0=#1\noindent\loop\sqr
  \advance\count0 by-1 \ifnum\count0>0\repeat}}
\def\tableau#1{\vcenter{\offinterlineskip
  \tableaustep=\tableauside\advance\tableaustep by-\tableaurule
  \kern\normallineskip\hbox
    {\kern\normallineskip\vbox
      {\gettableau#1 0 }%
     \kern\normallineskip\kern\tableaurule}%
  \kern\normallineskip\kern\tableaurule}}
\def\gettableau#1 {\ifnum#1=0\let\next=\null\else
  \squares{#1}\let\next=\gettableau\fi\next}

\begin{document}

\title{Two-level systems coupled to an oscillator: Excitation transfer and energy exchange }

\author{Peter L. Hagelstein}
\email{plh@mit.edu}
\author{Irfan U. Chaudhary}
\email{irfanc@mit.edu}
\affiliation{Research Laboratory of Electronics \\ 
Massachusetts Institute of Technology \\
Cambridge, MA 02139,USA}

\begin{abstract}
We consider models in which two sets of matched two-level systems
are coupled to a common oscillator in the case where the oscillator
energy is small relative to the two-level transition energies.
Since the two sets of two-level systems are coupled indirectly through
the oscillator, excitation transfer from one set of two-level systems
to the other is possible.
In addition, the excitation energy from the two-level systems may be
exchanged with the oscillator coherently, even though the oscillator
energy may be orders of magnitude smaller than the two-level system transition energy.
In the lossless case, we demonstrate these effects numerically, and
also use an approximate diagonalization to show that these effects
are expected from the model Hamiltonian.

We augment the model to include loss effects, and show that loss enhances
the excitation transfer effect by breaking the severe cancelation between
different paths that occurs in the lossless case.
We describe a simple approximate model wavefunction appropriate when the loss
increases rapidly with energy.
Within this model approximation, we present numerical and analytical results 
for excitation transfer and energy transfer rates, showing that
they are greatly increased.

Our study of these models is motivated in part by claims of excess heat production
in electrochemical experiments in heavy water.
We examine the question of whether the rates associated with this kind of
model are sufficiently large to be relevant to the experimental claims.
We find that consistency is possible given recent experimental results
showing strong screening effects in low energy deuteron-deuteron fusion
experiments in metals.

\end{abstract}

\maketitle

\section{Introduction}

   The Department of Energy conducted a review of cold fusion in 2004 
that included excess heat experiments as the primary focus of the written
review material and oral presentations.\cite{ReviewDoc} 
Although the reviewers were not asked to respond on the specific issue of
excess power (which has been the main goal of most of the experimental 
efforts), most of the reviewers volunteered comments on this aspect of
the review material.  
A majority of these comments were favorable.\cite{reviewcomments}
In addition, most of the reviewers recommended that research in this area be funded, 
and there was strong encouragement that those working in the field
should present their results in the mainstream scientific journals.

In successful experiments of this type, deuterium is introduced into a sample,
the sample is stimulated, and a large amount of energy is observed as thermal output.  
There are essentially no energetic nuclear products (such as neutrons, gammas, alphas, betas,
or x-rays that would signify the presence of energetic charged particles) seen in quantitative
correlation with the energy.
There is evidence for the production of helium in an amount quantitatively correlated
with the excess energy, such that an energy of about 24 MeV is measured in association
with each $^4$He atom detected.
There is no evidence that this helium was created with any significant kinetic energy.
A review of these issues and some discussion of the associated experiments is
given in the document prepared for the 2004 DoE review.\cite{ReviewDoc}

   With this as motivation, we are interested here in possible physical
mechanisms that may be involved, under the assumption that the excess heat effect
is real.
When two deuterons react in vacuum, the primary reaction channels are n+$^3$He and p+t,
with the reaction energy expressed as relative kinetic energy of the products, consistent
with our notions of energy and momentum conservation.
However, we know that the energy in these experiments is not due to the vacuum reaction pathway, since
the vacuum reaction products are not present in quantitative amounts. 
We would also not expect that these reactions should be responsible for the energy observed, 
since from vacuum nuclear physics the associated reaction rates are quite small.

   Whatever processes are responsible for the excess heat effect, they have not been
seen before in nuclear physics or condensed matter physics.
According to the 1989 DoE ERAB report\cite{ERAB}:
    ``Nuclear fusion at room temperature, of the type discussed
in this report, would be contrary to all understanding gained of nuclear reactions in 
the last half century; it would require the invention of an entirely new nuclear process.''
In this work, we consider models which correspond to proposed new physical processes.


\subsection{Basic scheme}
\epsfxsize = 4.00in
\epsfysize = 2.50in
\begin{figure} [t]
\begin{center}
\mbox{\epsfbox{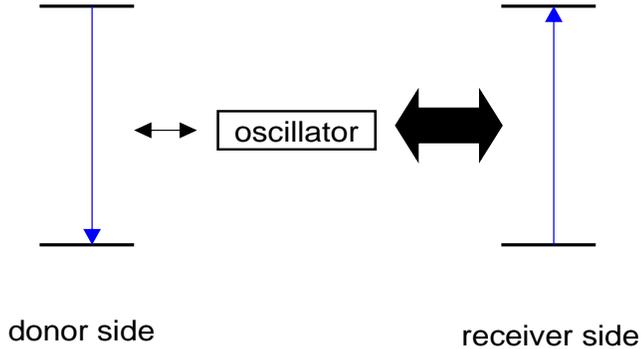}}
\caption{Schematic of two sets of two-level systems coupled to a common highly excited
oscillator.  The coupling between the donor side two-level systems and the oscillator is assumed
to be weak, as indicated by a small arrow between them.  The coupling between 
the receiver side two-level systems and oscillator is assumed to be strong,
as indicated by a large arrow between them.
}
\label{schemea}
\end{center}
\end{figure}

  For the past several years we have been studying a new class of candidate reaction
mechanisms to account for the excess heat effect.
At the heart of these physical mechanisms is the basic scheme illustrated in Figure \ref{schemea}.
In this figure, we show two sets of two-level systems which are individually coupled to a common oscillator.
On the donor side, the upper level is associated with molecular D$_2$, and the lower level is
associated with $^4$He.
On the receiver side, the lower level is associated with a ground state receiver nucleus, such as
a Pd isotope.
The upper level is associated with a (reasonably stable) excited state in the receiver nucleus.
Operation of the scheme involves the transfer of excitation from the two-level systems on
the donor side to the two-level systems on the receiver side, and subsequent loss of excitation
on the receiver side.
The coupling on the donor side is assumed to be weak, since the two deuterons must tunnel through
a Coulomb barrier in order to interact.
The coupling on the receiver side is assumed to be strong, since there is no requirement for tunneling
through a large barrier.

\subsection{Excitation transfer and energy exchange effects}

   From the literature on the closely-related spin-Boson problem,\cite{Cibils1991}    
one would expect interesting (fast) oscillatory dynamics in this kind of model to occur 
near the oscillator frequency, and also near frequencies associated with the two-level transitions energies.
Such effects are not of interest to us in this work.
Instead, we are concerned with two different effects (excitation
transfer and energy exchange) which occur on a slower timescale (in the absence of loss).
%
%
The excitation transfer effect (as we use the term in this paper) involves moving excitation from one
set of two-level systems to the other, through indirect coupling with the oscillator.
The energy exchange effect (as we use the term in this paper) involves an increase (or decrease) of
oscillator energy in association with a corresponding decrease (or increase) in the number of excited
two-level systems.
These effects are weak in the absence of loss, and it is only specially constructed solutions
for models where resonance conditions are precisely matched that show these effects clearly.
The possibility that energy can be exchanged coherently between two quantum systems with incommensurate
energies is not widely appreciated, and one of the goals of this manuscript is to draw
attention to it through models, analytic results, and numerical calculations.
The analysis of this manuscript focuses on the coupling with a single highly-excited phonon mode
(the oscillator in Figure \ref{schemea}), hence coupling to other modes is not a
focus of the models.  As a result, we do not obtain here an estimate for loss of two-level 
excitation through sequential incoherent phonon exchange.

Excitation transfer as a quantum effect is well known;
in biophysics and nanoscale physics, one encounters F\"orster excitation
transfer, in which Coulombic dipole coupling between molecules allows excitation at
one location to be transferred to another location.\cite{Forster,RET1999}
There is no possibility of a F\"orster type of (Coulomb-mediated) excitation transfer between nuclei
at the MeV energy scale analogous to the effect in atoms and molecules. 
Instead, we focus on a second-order indirect version of the effect.
This kind of scheme has been considered previously in association with 
the possibility of indirect excitation transfer of optical excitation
through coupling with a common microwave mode.\cite{XRL}
Here, excited nuclear states are assumed to couple to a common phonon mode, with a resulting
indirect second-order coupling to excited nuclear states at other sites. 
Such indirect coupling has the potential to support a site-to-other-site excitation
transfer effect.
An indirect excitation transfer effect of this kind should in principle be possible
when two atoms or molecules are coupled to a common phonon mode.
However, we are not aware of this kind of indirect second-order process having 
yet produced an important contribution in an experiment.
Second-order indirect coupling itself is discussed widely in the literature.\cite{Juzel1994}
Indirect coupling between quantum well states interacting with a common phonon mode
has been discussed in the literature.\cite{Vorrath2003}


\subsection{Overview of the paper}

  The number of issues involved in this scheme is very large, and we
have no hope of addressing them all here.
Instead, we focus attention on basic issues associated with the idealized
model illustrated in Figure \ref{schemea},
to develop understanding of the excitation transfer and energy exchange
mechanisms.
The layout of the paper is as follows.
We begin in Section II with a discussion of the basic model for two sets of two-level systems
coupled to a common oscillator.  
We present numerical results which illustrate the
presence of an excitation transfer effect and energy exchange effect.  
Since the idealized model is relatively simple, we are able to apply a rotation that
brings out terms which exhibit these effects individually.
Both effects in this model are quite weak, and require precise resonances to be observed.


  In Section III, we augment the model to include loss.
The specific loss mechanisms that are most relevant to the
discussion are those in which a unit of two-level transition
energy is lost.
This can occur through the decay of excited states, 
or through more subtle processes in which the coupled system
is capable of dissipating a large quantum of energy through
a variety of decay modes.
These latter processes have a big impact on the quantum states
and associated dynamics, 
and are of much interest to us in this paper
(we assume that the physical states associated with the two-level 
systems are reasonably stable at their given energies). 
Loss mechanisms are discussed in Appendix A.
The impact of loss on excitation transfer is examined using perturbation theory. 
The associated indirect coupling in the lossless model is weak due to
destructive interference between different pathways.
This destructive interference is removed when loss is included, which
leads to a dramatic increase in the indirect interaction.


  In Section IV, we focus on energy exchange between the two-level systems on
the receiver side and the oscillator in the presence of loss (at the two-level transition
energy).
To make quantitative estimates, we require a specification of the loss terms
in the Hamiltonian.
The inclusion of loss results in a dramatic increase in rates over the lossless
case as mentioned above; however, different loss models produce only minor changes
in the increased rates.
In Appendix B we discuss an approximate wavefunction in which parts which see
significant loss are omitted.
Such an approximate wavefunction would result from a loss model in which the
loss increases strongly with energy.
The advantage of this kind of approximation is that 
it leads to a well-defined and simple approximate wavefunction 
which can be used to evaluate lossy models systematically,
and that it brings out the dominant effects associated with the inclusion of loss
in the model.
We make use of this approximation in order to develop numerical results for energy
exchange in the case of intermediate coupling, and analytic results (Appendix C) in the
case of strong coupling.
%
%
In Section V, we focus on excitation transfer between donor and receiver two-level
systems.
We again make use of the approximate wavefunction discussed in Appendix B for
a direct numerical calculation of excitation transfer in the case of intermediate coupling,
and analytic results (Appendix D) in the case of strong coupling.

In Section VI, we revisit the issue of excitation transfer using a different
approach.
In the limit that the number of oscillator quanta is very large and many two-level
systems are involved, then we may model the system locally as being periodic.
Calculations of the associated band structure leads to estimates of the group
velocity associated with excitation transfer which compare well with the
analytic results obtained in Appendix D.
One new feature that results from this model is that  
we find that excitation transfer can proceed at nearly the maximum
rate possible, even in the absence of a precise resonance when the receiver-side
two-level systems are strongly coupled to the oscillator.
This very interesting result can be understood as involving a coupling of
energy directly with the oscillator to make up the power loss or gain
associated with mismatched excitation transfer.


In Section VII, we consider the excitation transfer rate (assuming it is rate
limiting) in comparison with excess power production in excess heat experiments.
It is not possible to obtain consistency without the inclusion of screening
effects.
Experimental results indicate the presence of strong screening at low energies;
however, a quantitative explanation for the effect from theory is currently lacking.
In light of this situation, we conclude that the model results appear
to be consistent with excess heat given what seems to be a reasonable
value for the (zero-relative energy) screening energy.


In Section VII, we summarize our results and provide some discussion of the
significance.

\newpage
\section{Sets of two-level systems coupled to an oscillator}

   Our first goal is to present a highly idealized model which illustrates
the excitation transfer and energy transfer effects.
We study an idealized Hamiltonian that describes two sets of two-level systems coupled
to an oscillator.
Such a model exhibits both excitation transfer and energy transfer effects, as
can be shown by a direct numerical solution.
Because the Hamiltonian is relatively simple, we are able to develop an approximate diagonalization.
The result of this approximate diagonalization is the development of an interaction term that 
corresponds to the excitation transfer effect discussed above, as well as an interaction
term that can couple energy between the receiver-side two-level systems and the oscillator.

\subsection{Idealized model}

The associated Hamiltonian is

{\small

\begin{equation}
\hat{H}
~=~
\Delta E_1  {\hat{S}_z^{(1)} \over \hbar}
+
\Delta E_2  {\hat{S}_z^{(2)} \over \hbar}
+
\hbar \omega_0 \hat{a}^\dagger \hat{a}
+
V_1 e^{-G} {2 \hat{S}_x^{(1)} \over \hbar}(\hat{a}+\hat{a}^\dagger)
+
V_2 {2 \hat{S}_x^{(2)} \over \hbar}(\hat{a}+\hat{a}^\dagger)
\label{lossless}
\end{equation}

}

\noindent
where the oscillator energy is presumed to be much less than the transition energy of the two-level
systems

\begin{equation}
\hbar \omega_0 \ll \Delta E_1
\ \ \ \ \ \ \ \ \ \
\hbar \omega_0 \ll \Delta E_2
\end{equation}

\noindent
We see in this model a set of two-level systems with transition energy $\Delta E_1$
which are weakly coupled to the oscillator.  
We use a pseudospin $\hat{S}_z$ operator to keep track of how many of the two-level 
systems are excited.  
Transitions from the upper state to the lower state are described using the pseudospin $\hat{S}_x$
operator, which assumes that the coupling is independent of which two-level system is involved,
and that we would expect Dicke enhancement factors\cite{Dicke} to be present.
The strength of this coupling is $V_1 e^{-G}$, where the presence of the Gamow factor $e^{-G}$
signals that this transition is hindered.
We have assumed linear coupling with the oscillator for simplicity, although our basic results
would change little had we instead used a weak nonlinearity.
We see a second set of two-level systems with transition energy $\Delta E_2$ which
are more strongly coupled to the oscillator.
We assume that the oscillator is linear.

\subsection{Numerical solution exhibiting excitation transfer effect}
\epsfxsize = 3.80in
\epsfysize = 2.80in
\begin{figure} [t]
\begin{center}
\mbox{\epsfbox{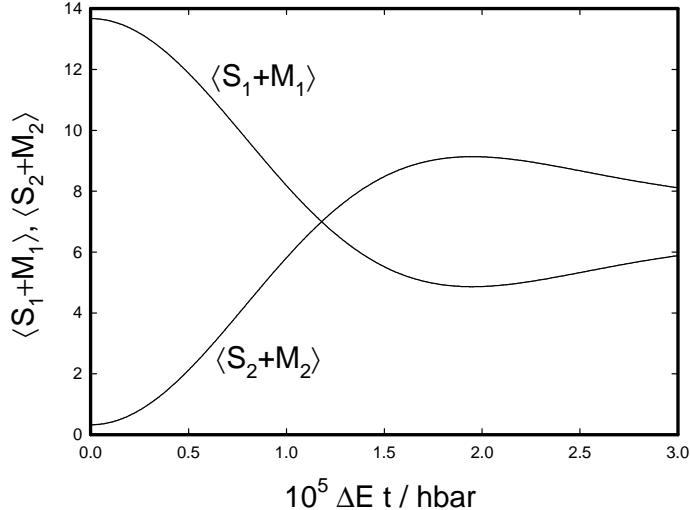}}
\caption{Demonstration of excitation transfer in the loss-free system.
Expectation values of the number of excited two-level systems from a numerical
calculation with one set of 14 two-level systems transferring excitation to
a second set of 14 two-level systems.  In this calculation, we used 
 $V_2 \sqrt{n_0}/\Delta E = 0.1$.  Since excitation transfer requires
precise resonances, we chose $V_1e^{-G} \sqrt{n_0}/\Delta E=0.1$ in order to maximize
the total amount of excitation transfer in this example.  We used a window
of oscillator quanta with $n_0 \le n \le n_0+21$, where the oscillator 
energy is 0.03 $\Delta E$.  A time-dependent wavefunction was constructed of the 15 nearly
resonant eigenfunctions, and initialized to maximize the probability of
occupation of the state with 14 excited two-level systems in the first set,
no excited two-level systems in the second set, and 10 oscillator quanta.}
\label{nvst14}
\end{center}
\end{figure}

   We consider now two numerical solutions to the Schr\"odinger equation to 
illustrate the effects under discussion.
We show first the excitation transfer effect, in which the excitation of two-level
systems in the first system are transferred to the second-system.
The result of a calculation is presented in Figure \ref{nvst14} 
which shows excitation transfer from one set of 14 
two-levels systems to a second set of 14 two-level systems.  
Excitation transfer in this idealized model is a slow process that depends
critically on the energies being resonant.
Since coupling with the oscillator produces a self-energy shift, the most
dramatic effect can be obtained under conditions where the two sets of
two-level systems are matched in number and coupling strength.
Consequently, we have chosen to illustrate this effect here with a matched
system (even though we are interested in the model when the coupling
on the donor side is hindered).
In addition, the strongest indirect coupling effect is developed when the
number of oscillator quanta is not so great, since the relative strength of 
indirect coupling in this model is reduced by $O(n)$ (due to severe
cancelation of terms from different paths) as compared to the 
first-order interaction.
The numerical calculation in this case illustrates that excitation transfer is
present in this model, and that the associated rate is slow.

\subsection{Numerical solution exhibiting energy exchange effect}


\epsfxsize = 3.80in
\epsfysize = 2.80in
\begin{figure} [t]
\begin{center}
\mbox{\epsfbox{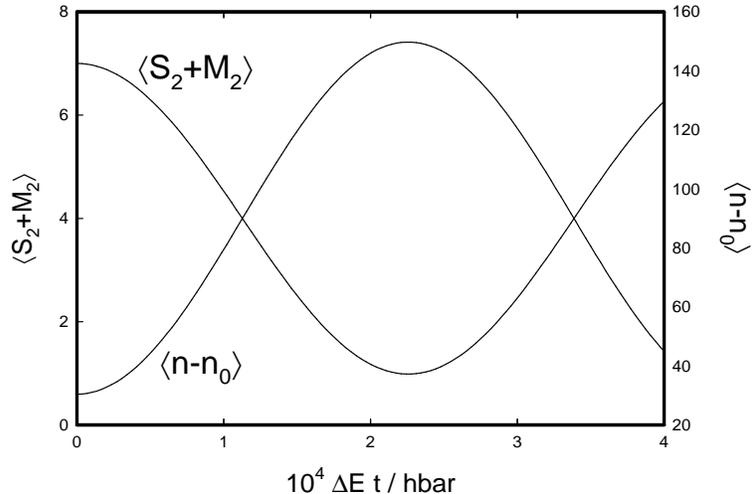}}
\caption{Demonstration of energy exchange in the loss-free system.
Expectation values of the number of excited two-level systems from a numerical
calculation with 8 receiver-side two-level systems coupled to an oscillator.
For this calculation we used an oscillator energy  of $\hbar \omega_0 = \Delta E_2/11$, 
and a coupling strength of  $V_2 \sqrt{n_0}/\Delta E_2 = 0.33744$.  This coupling
strength is selected so that the dressed receiver-side transition energy 
approximately matches 15 oscillator quanta (since the dressed states are made up
of a superposition of different $M_2$ states, the mixed states occur at
roughly 0.76 $M_2$, so that roughly 20 oscillator quanta make up one bare
two-level system quantum).  
For simplicity, we assumed no coupling with the first set of two-level systems.
The total number of oscillator
quanta in this calculation is chosen to be $10^8$.  In this calculation, 
approximately resonant eigenfunctions were combined to make a time-dependent
solution that is initialized with the receiver-side two-level systems nearly
completely excited.  One sees that the energy of the two-level systems is
transferred slowly to the oscillator and back.}
\label{exch8}
\end{center}
\end{figure}


We next examine energy exchange on the receiver side.
To illustrate this effect, we consider a model in which the hindered coupling with
the first set of two-level systems is taken to be zero, which isolates the coupling
between the two-level systems and oscillator on the receiver side.
The most dramatic effect is seen under conditions where the dressed two-level system
transition energy becomes approximately independent of oscillator excitation, which
occurs for specific solutions at large oscillator quantum number $n$.
(The bare receiver-side transition energy is $\Delta E_2$, which shifts to 
higher energy in the presence of coupling; we refer to the transition
energy shifted in this way as the dressed transition energy.)
When the receiver-side coupling is made stronger then more oscillator
quanta can be exchanged for a two-level system transition energy.
In addition, there needs to be a precise resonance between the dressed two-level
system energy and an odd multiple of the oscillator energy (since the Hamiltonian 
does not mix states with even and odd combinations of oscillator quanta and two-level
system excitation).
The results of a numerical calculation are presented in Figure \ref{exch8} in which
eight two-level systems initially in the ground state become excited, with a 
corresponding loss of energy present in the highly excited oscillator.
Although the coupling of the two-level systems with the oscillator is linear, in this
calculation approximately 15 oscillator quanta are matched in energy to a single
dressed two-level system transition energy.
One sees from this numerical calculation that energy can be transferred from the
oscillator to the two-level systems, and that the associated energy transfer
rate is slow.

\subsection{Rotation}

   We have studied this system with a variety of tools, both analytic and
numerical.  The results of these studies can be understood most simply in terms of
an approximate diagonalization that we can implement with a rotation

\begin{equation}
\hat{H}^\prime 
~=~ 
\hat{U} \hat{H} \hat{U}^\dagger
\end{equation}

\noindent
using the unitary operator $\hat{U}$ defined by

{\small

\begin{equation}
\hat{U} 
~=~ 
\exp 
\left \lbrace
i \arctan
\left [
{2V_1 e^{-G} (\hat{a}+\hat{a}^\dagger) \over \Delta E_1}
\right ]
{\hat{S}^{(1)}_y \over \hbar}
+
i \arctan
\left [
{2V_2 (\hat{a}+\hat{a}^\dagger) \over \Delta E_2}
\right ]
{\hat{S}^{(2)}_y \over \hbar}
\right \rbrace
\end{equation}

}

\noindent
This rotation is similar to the single spin version of the
rotation given by Wagner.\cite{Wagner}
The rotation can be carried out exactly to produce

\begin{samepage}

{\small

$$
\hat{H}^\prime
~=~
\sqrt{ \Delta E_1^2 + [2V_1 e^{-G} (\hat{a}+\hat{a}^\dagger)]^2 }
{\hat{S}^{(1)}_z \over \hbar}
~~+~~
\sqrt{ \Delta E_2^2 + [2V_2 (\hat{a}+\hat{a}^\dagger)]^2 }
{\hat{S}^{(2)}_z \over \hbar}
~~+~~
\hbar \omega_0 \hat{a}^\dagger \hat{a}
\ \ \ \ \ \ \ \ \ \ \ \ \ \ \ \ \ \ \ \ \ \ \ \ \ \ \ \ \ \ \ \ \ \ \ \
\ \ \ \ \ \ \ \ \ \ \ \ \ \ \ \ \ \ \ \ \ \ \ \ \ \ \ \ \ \ \ \ \ \ \ \
$$
$$
~+~
i {\hbar \omega_0 \over 2}
\left \lbrace
{
\left [ 
\displaystyle {V_1 e^{-G} \over \Delta E_1 }
\right ]
\over 
\left [ 1 + \displaystyle{4V_1^2 e^{-2G}(\hat{a}+\hat{a}^\dagger)^2 \over \Delta E_1^2} \right ]
}
(\hat{a}-\hat{a}^\dagger)
+
(\hat{a}-\hat{a}^\dagger)
{
\left [ 
\displaystyle {V_1 e^{-G} \over \Delta E_1 }
\right ]
\over 
\left [ 1 + \displaystyle{4V_1^2 e^{-2G}(\hat{a}+\hat{a}^\dagger)^2 \over \Delta E_1^2} \right ]
}
\right \rbrace
{2\hat{S}^{(1)}_y \over \hbar}
$$
$$
~+~
i {\hbar \omega_0 \over 2}
\left \lbrace
{
\left [ 
\displaystyle {V_2 \over \Delta E_2 }
\right ]
\over 
\left [ 1 + \displaystyle{4V_2^2(\hat{a}+\hat{a}^\dagger)^2 \over \Delta E_2^2} \right ]
}
(\hat{a}-\hat{a}^\dagger)
+
(\hat{a}-\hat{a}^\dagger)
{
\left [ 
\displaystyle {V_2 \over \Delta E_2 }
\right ]
\over 
\left [ 1 + \displaystyle{4V_2^2(\hat{a}+\hat{a}^\dagger)^2 \over \Delta E_2^2} \right ]
}
\right \rbrace
{2\hat{S}^{(2)}_y \over \hbar}
$$
$$
~+~
\hbar \omega_0 
{
\left [ 
\displaystyle {V_1 e^{-G} \over \Delta E_1 }
\right ]^2
\over 
\left [ 1 + \displaystyle{4V_1^2 e^{-2G}(\hat{a}+\hat{a}^\dagger)^2 \over \Delta E_1^2} \right ]^2
}
\left ( {2 \hat{S}^{(1)}_y \over \hbar} \right )^2
~+~
\hbar \omega_0 
{
\left [ 
\displaystyle {V_2 \over \Delta E_2 }
\right ]^2
\over 
\left [ 1 + \displaystyle{4V_2^2(\hat{a}+\hat{a}^\dagger)^2 \over \Delta E_2^2} \right ]^2
}
\left ( {2 \hat{S}^{(2)}_y \over \hbar} \right )^2
$$
\begin{equation}
~+~
2\hbar \omega_0 
{
\left [ 
\displaystyle {V_1 e^{-G} \over \Delta E_1 }
\right ]
\over 
\left [ 1 + \displaystyle{4V_1^2 e^{-2G}(\hat{a}+\hat{a}^\dagger)^2 \over \Delta E_1^2} \right ]
}
{
\left [ 
\displaystyle {V_2 \over \Delta E_2 }
\right ]
\over 
\left [ 1 + \displaystyle{4V_2^2(\hat{a}+\hat{a}^\dagger)^2 \over \Delta E_2^2} \right ]
}
\left ( {2 \hat{S}^{(1)}_y \over \hbar} \right )
\left ( {2 \hat{S}^{(2)}_y \over \hbar} \right )
\end{equation}

}

\end{samepage}

\vskip 0.20in

\noindent
Although this rotated Hamiltonian looks complicated, it illustrates the effects under discussion.
The rotation has eliminated the first-order coupling terms that appeared
in the initial Hamiltonian [Equation (\ref{lossless})], 
and replaced them with much smaller first-order and second-order coupling terms.  
The dressed two-level systems are now nearly decoupled, with
energies that are dependent on the excitation of the oscillator.

In this rotated Hamiltonian there occurs a weak second-order term that 
mediates transitions between the two sets of two-level systems:

{\small

$$
2\hbar \omega_0 
{
\left [ 
\displaystyle {V_1 e^{-G} \over \Delta E_1 }
\right ]
\over 
\left [ 1 + \displaystyle{4V_1^2 e^{-2G}(\hat{a}+\hat{a}^\dagger)^2 \over \Delta E_1^2} \right ]
}
{
\left [ 
\displaystyle {V_2 \over \Delta E_2 }
\right ]
\over 
\left [ 1 + \displaystyle{4V_2^2(\hat{a}+\hat{a}^\dagger)^2 \over \Delta E_2^2} \right ]
}
\left ( {2 \hat{S}^{(1)}_y \over \hbar} \right )
\left ( {2 \hat{S}^{(2)}_y \over \hbar} \right )
$$

}

\noindent
This term is associated with the excitation transfer effect that we discussed above, and which
was illustrated in our first numerical example.

Also present is a weak second-order interaction term which mediates the energy exchange
effect:

{\small

$$
\hbar \omega_0 
{
\left [ 
\displaystyle {V_2 \over \Delta E_2 }
\right ]^2
\over 
\left [ 1 + \displaystyle{4V_2^2(\hat{a}+\hat{a}^\dagger)^2 \over \Delta E_2^2} \right ]^2
}
\left ( {2 \hat{S}^{(2)}_y \over \hbar} \right )^2
$$

}

\noindent
In the rotated problem, there occurs a direct coupling between basis states that differ by
two units of excitation in the receiver-side two-level systems and a modest number of
oscillator quanta.
This term illustrates the energy transfer effect that we illustrated in our
second numerical example.
There are in addition weak residual first-order terms that are capable of causing single two-level
excitation and de-excitation also with the exchange of a moderate number of oscillator quanta.
%
%

This provides a confirmation that the functionality that we are looking for is
present in a very simple model for the coupling of two sets of two-level 
systems with an oscillator.  
We have demonstrated an excitation transfer
effect, and we have also demonstrated an energy exchange effect.
However, in this model both effects are weak and require us to maintain resonances
in order to see the effects in calculations.

\newpage

\section{Inclusion of loss}

  New decay pathways may be present in a physical system where two-level
systems are coupled to an oscillator.
For example, suppose that the oscillator under discussion is a highly
excited phonon mode in a lattice, and that the states of the two-level
systems correspond to nuclear states that differ by MeV.
A coupling between the two systems will inevitably produce a coupling
to states in which rapid decay processes are energetically allowed.

\begin{figure} [t]
\begin{center}
\mbox{\epsfbox{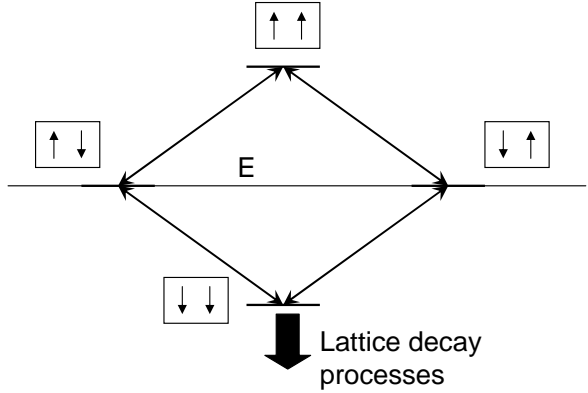}}
\caption{Schematic of levels for a pair of two-level systems with indirect coupling
between two degenerate states, augmented with loss.  Coupling
to the lower state with zero units of excitation leads to a situation in
which the state energy (which is close to that of a basis state with one unit
of two-level excitation) is much greater than the basis state energy.  In the
model with loss under discussion, the lattice has the possibility of expressing
the energy difference through loss channels.  The destructive interference
between the upper and lower pathway is now broken.
}
\label{levels2}
\end{center}
\end{figure}

For example, suppose that only two two-level systems are present,
and one is initially excited.
First-order coupling will produce a mixing with states in which both
two-level systems are excited, and in which both two-level systems
are in the ground state 
(second-order indirect coupling is viewed similarly in other problems\cite{Jenkins}). 
%
%
In the latter case, the two-level systems and oscillator will be in a
virtual state in which there is an energy surplus.
One would expect that the lattice would very quickly find decay modes
for this energy excess 
(this mechanism was recognized several years ago\cite{ICCF7} and
remains a candidate to account for anomalous alpha emission\cite{APS2000}).
This is illustrated schematically in Figure \ref{levels2}.

In the idealized model considered in the previous section 
there is no consequence associated with coupling to such states.  
The primary advantages of this kind of model to us are that the coupled oscillator and 
two-level system has been studied in the literature, 
and that we are able to develop an approximate diagonalization
which makes clear that both excitation transfer and energy
transfer effects are present.
However, to develop more realistic models we will need to augment the idealized
Hamiltonian of Equation (\ref{lossless}) with loss terms that will include these
lattice decay pathways.
The inclusion of such terms changes the problem significantly, as we discuss below.

\subsection{Idealized model augmented with loss}




\epsfxsize = 3.20in
\epsfysize = 2.00in


  To augment the model with loss, we adopt a description based on a sector decomposition of the
available state space.  
Processes that preserve sector are described with sector-specific operators that are Hermitian.  
Processes that remove probability amplitude from one sector to another are described 
by operators that are not Hermitian with respect to a single sector, but are Hermitian 
over all sectors.  
Suppose that we assume that lattice-induced decay processes near 24 MeV involve energetic
decay products (such as alpha particles, protons, neutrons, etc.).  
We could then define our sector of interest as one which lacks any such energetic decay products.  
We employ an infinite-order Brillouin-Wigner approach to eliminate the sector with energetic decay products, 
and the loss is accounted for through a non-Hermitian term in the sector Hamiltonian
for our sector of interest.
The approach is equivalent to that discussed by Feshbach,\cite{Feshbach} although our application
and language are different.

The idealized Hamiltonian is augmented with loss by first restricting it to the sector of interest, 
and then including decay processes using a loss operator that is anti-Hermitian with respect to the sector.  
This produces

{\small

\begin{equation}
\hat{H}
~=~
\Delta E_1  {\hat{S}_z^{(1)} \over \hbar}
+
\Delta E_2  {\hat{S}_z^{(2)} \over \hbar}
+
\hbar \omega_0 \hat{a}^\dagger \hat{a}
-
{i \hbar \over 2}\hat{\Gamma}(E)
+
V_1 e^{-G} {2 \hat{S}_x^{(1)} \over \hbar}(\hat{a}+\hat{a}^\dagger)
+
V_2 {2 \hat{S}_x^{(2)} \over \hbar}(\hat{a}+\hat{a}^\dagger)
\label{lossy}
\end{equation}

}

\noindent
The loss term in this model may be thought of as arising from second-order terms of
the form

$$
i~
{\rm Im}
\bigg \lbrace
\hat{V} [E - \hat{H}_0]^{-1} \hat{V}
\bigg \rbrace
$$

\noindent
which are explicitly dependent on the energy $E$.
Loss mechanisms and operators are discussed briefly in Appendix A.

\subsection{Excitation transfer in weak coupling}

  We turn our attention first to the issue of the existence of an excitation transfer
effect in this augmented model.
Although we are most interested in model systematics in the strong coupling limit, it
is convenient here to consider the excitation transfer effect in the weak coupling
limit using second-order perturbation theory.
The relevant second-order interaction is derived from

{\small

$$
V_1 e^{-G} {2 \hat{S}_x^{(1)} \over \hbar}(\hat{a}+\hat{a}^\dagger)
[E-\hat{H}_0]^{-1}
V_2 {2 \hat{S}_x^{(2)} \over \hbar}(\hat{a}+\hat{a}^\dagger)
~+~
V_2 {2 \hat{S}_x^{(2)} \over \hbar}(\hat{a}+\hat{a}^\dagger)
[E-\hat{H}_0]^{-1}
V_1 e^{-G} {2 \hat{S}_x^{(1)} \over \hbar}(\hat{a}+\hat{a}^\dagger)
$$

}

\noindent
If we keep near-resonant terms, then we obtain

{\small

$$
\hat{V}_{12} 
~=~
V_1 V_2 e^{-G}
{\hat{S}_+^{(1)} \over \hbar}
{\hat{S}_-^{(2)} \over \hbar}
\bigg (
  {\hat{a}^\dagger \hat{a}   \over \Delta E_1 + \hbar \omega_0 + i \hbar \Gamma/2}
+ {\hat{a}\hat{a}^\dagger    \over \Delta E_1 - \hbar \omega_0 + i \hbar \Gamma/2}
- {\hat{a}^\dagger \hat{a}   \over \Delta E_2 - \hbar \omega_0}
- {\hat{a}\hat{a}^\dagger    \over \Delta E_2 + \hbar \omega_0}
\bigg )
$$
$$
~+~
V_1 V_2 e^{-G}
{\hat{S}_-^{(1)} \over \hbar}
{\hat{S}_+^{(2)} \over \hbar}
\bigg (
  {\hat{a}^\dagger \hat{a}   \over \Delta E_1 + \hbar \omega_0 + i \hbar \Gamma/2}
+ {\hat{a}\hat{a}^\dagger    \over \Delta E_1 - \hbar \omega_0 + i \hbar \Gamma/2}
- {\hat{a}^\dagger \hat{a}   \over \Delta E_2 - \hbar \omega_0}
- {\hat{a}\hat{a}^\dagger    \over \Delta E_2 + \hbar \omega_0}
\bigg )
$$

\noindent
In this equation, the loss term $\Gamma(E)$ is to be evaluated with a surplus of
roughly $\Delta E_1$.

If we were to set these loss terms to zero, then we could compare this interaction with what we
obtained previously.  
This is most interesting and relevant on resonance when $\Delta E_1=\Delta E_2$.
In this case the second-order interaction is approximately

\begin{equation}
\hat{V}_{12} 
~\rightarrow~
{
2 \hbar \omega_0 V_1 V_2 e^{-G}
\over 
\Delta E^2}
\left [
{\hat{S}_+^{(1)} \over \hbar}
{\hat{S}_-^{(2)} \over \hbar}
+
{\hat{S}_-^{(1)} \over \hbar}
{\hat{S}_+^{(2)} \over \hbar}
\right ]
\ \ \ \ \ \ \ \ \ \
[\Gamma(E) \rightarrow 0]
\end{equation}

\noindent
This is in agreement with the results obtained using our approximate diagonalization in
the previous section.

On the other hand, if the decay rate is enormous we
would obtain approximately

\begin{equation}
\hat{V}_{12} 
~\rightarrow~
-
{
2  V_1 V_2 e^{-G}
\over 
\Delta E}
( \hat{a}^\dagger \hat{a} + \hat{a}\hat{a}^\dagger )
\left [
{\hat{S}_+^{(1)} \over \hbar}
{\hat{S}_-^{(2)} \over \hbar}
+
{\hat{S}_-^{(1)} \over \hbar}
{\hat{S}_+^{(2)} \over \hbar}
\right ]
\ \ \ \ \ \ \ \ \ \
[\Gamma(E) \rightarrow \infty]
\end{equation}

\noindent
The effect of loss in this case is to destroy destructive interference
between the different pathways, which leads to a dramatic enhancement of
the excitation transfer rate.

\newpage

\section{Energy exchange with the oscillator}

  We now turn our attention to the energy exchange effect when loss (at
the two-level system transition energy) is present. 
In our study of the lossless version of the problem, we observed that after rotation
a self-energy term was present that produced an energy shift, and also an
additional term was present that could exchange two units of excitation in
association with the transfer of a modest number of oscillator quanta.
Based on the above perturbation theory results, in the presence of loss we would
expect a much larger self-energy shift and energy exchange effect (even though
we do not at present have a simple rotation to bring out the effects so
cleanly).
We consider in this section first a numerical result in intermediate coupling,
and then an analytic result in strong coupling, to illustrate this.

\subsection{Receiver-side self-energy and energy exchange: Intermediate coupling strength}

   To develop numerical solutions for the lossy case, we require a specific loss model. 
We have studied energy exchange with a number of different loss models, with the result
that once the destructive interference is broken as discussed above, then the self-energy
and energy exchange effects are increased dramatically as compared to the lossless case.
Relatively minor differences in comparison occur between the results with different loss models.
In Appendix B we describe an approximate wavefunction which captures the dominant
effect (the elimination of destructive interference from contributions from different
paths) by omitting parts of the wavefunction most impacted by loss in a model with
loss that increases strongly with energy.
Such a model results in a flattening of the wavefunction above a constant energy
line in $(n,M_2)$-space [or a surface in $(n,M_1,M_2)$-space]. 
[The notion of  $(n,M_1,M_2)$-space is connected with the construction of a solution using
$|n \rangle |S_1, M_1 \rangle | S_2, M_2 \rangle$ product basis states.]


\epsfxsize = 3.50in
\epsfysize = 3.00in
\begin{figure} [t]
\begin{center}
\mbox{\epsfbox{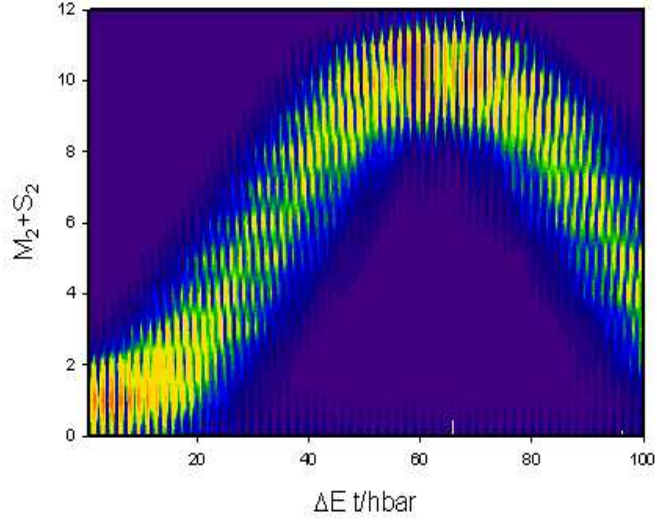}}
\caption{Demonstration of energy exchange between the receiver-side two-level systems 
and the oscillator in the case of infinitely fast loss.
In this calculation, twelve receiver-side two-level systems were included with the oscillator (the coupling with the first
set of two-level systems was set to zero).  
The coupling strength was taken to be $V_2 \sqrt{n_0}/\Delta E = 0.4$.
The oscillator energy was taken to be $\hbar \omega_0 = \Delta E_2/15$, 
and we assumed an oscillator excitation near $n_0 = 10^8$.
In this calculation, the full model sector Hamiltonian with loss was diagonalized,
and a dynamical solution was developed through a selection of components localized
near the twelve-oscillator ground state.
Plotted is the logarithm of the probability distribution for $M_2+S_2$, which is
proportional to the excitation of the receiver-side two-level systems, as a function of time.
One observes a very rapid and nearly complete coherent excitation of the two-level systems.
The associated probability distribution for the oscillator in this calculation is
illustrated in the following figure.}
\label{Exchvst3b}
\end{center}
\end{figure}


\epsfxsize = 3.50in
\epsfysize = 3.00in
\begin{figure} [t]
\begin{center}
\mbox{\epsfbox{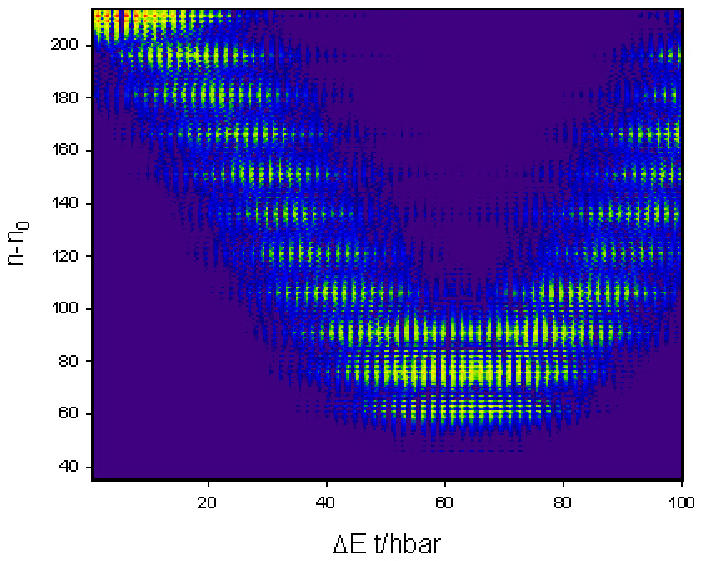}}
\caption{Logarithm of the probability distribution of the oscillator states as a function of time for the calculation
described in the previous figure.  
One sees in this calculation that the energy going into the excitation of the twelve
two-level systems is lost from the oscillator.}
\label{Exchvst3a}
\end{center}
\end{figure}


The results of a calculation based on this kind of approximation 
illustrating energy exchange between the oscillator and twelve two-level 
systems is shown in Figures \ref{Exchvst3b} and \ref{Exchvst3a}.  
In the first of these, the probability distribution of the receiver-side
two-level systems is shown as a function of time.  
One observes that energy is transferred to the two-level systems rapidly,
much faster (Figure \ref{Exchvst3b}) than in the lossless case that we considered above (Figure \ref{exch8}).
One observes that the energy increase associated with the excitation of the
two-level systems matches the energy loss from the oscillator
(Figure \ref{Exchvst3a}).

\subsection{Receiver-side self-energy and energy exchange: Strong coupling}

   In the limit of strong coupling between the oscillator and two-level systems 
on the receiver side, we are able to develop approximate analytic results. 
This is discussed in more detail in Appendix C.
This approximate model is consistent with the assumption that the oscillator energy
can be neglected (which makes the problem approximately periodic locally) to 
develop an estimate of the local self-energy.
The result of this analysis is an approximate self-energy given by

\begin{equation}
E_2(\theta) 
~\approx~
E_0
-
2 V_2 \sqrt{n_0} \cos(\theta) \sqrt{S_2^2 - (M_2)_0^2}
\label{enexch}
\end{equation}

\noindent
where $E_2$ refers to the energy of the oscillator and the receiver-side
two-level systems (the first set of two-level systems being omitted from
this calculation), where $E_0$ here is the bare energy around which the solution
is developed

\begin{equation}
E_0 ~=~ 
\left [
\Delta E_2 M_2
+
\hbar \omega_0 n
\right ]_{(n,M_2)_0}
\end{equation}

\noindent
The phase $\theta$ refers to the local phase in the $e^{i n \theta}$ dependence 
of the solution on $n$ assumed in the approximate local solution.
The variable $\theta$ is conjugate to $n$, and can be identified with a wave momentum
in the $n$-direction.
Associated with this self-energy estimate, we can develop
a group velocity estimate given by

\begin{equation}
{d n \over dt}
~\sim~
{1 \over \hbar} {d E_2 \over d \theta}
~=~
{2 V_2 \sqrt{n_0} \sin (\theta) \sqrt{S_2^2 - (M_2)_0^2} \over \hbar}
\end{equation}

\noindent
In essence, free energy exchange between the oscillator and two-level systems
occurs in the strong coupling regime [where the maximum rates occur for
$\sin(\theta) = \pm 1$].
This is consistent with the effect seen in the numerical result discussed above for energy
exchange in the case of intermediate coupling strength.
It is also consistent with more sophisticated results we have obtained 
using the method outlined in Section VI (but which are not presented explicitly in this paper).

\newpage

\section{Excitation transfer }

  In the weak coupling limit, we have seen that the second-order interaction
which is associated with excitation transfer is much increased due
to the presence of loss.
Once again, we require different tools to study the system when the coupling
on the receiver side is of intermediate strength or is strongly coupled (the
coupling from the first set of two-level systems in this model will always
be weak).

\subsection{Excitation transfer: Intermediate coupling strength}

We first consider numerical results for the case where the receiver-side
coupling is of intermediate strength.
In Figures \ref{Trans1} and \ref{Trans1a} we show numerical solutions 
(in the approximation described in Appendix B)
for the Schr\"odinger equation using the model Hamiltonian augmented with loss
[Equation (\ref{lossy})]  
which shows excitation transfer (Figure \ref{Trans1})
under conditions where the energy exchange with the oscillator is small
(Figure \ref{Trans1a}).
From this calculation, one sees that excitation is transferred from
the first set of two-level systems to the second set, with an associated
rate which is much faster (Figure \ref{Trans1})
than what we found previously in the lossless case (Figure \ref{nvst14}).
 
In the lossless case, we observed that it was important to match the 
dressed transition energies in order to maximize the effect.
Here, the self-energy shift on the receiver side is much greater, and
in general it is more difficult to match dressed transition energies.
Nevertheless, in the model with loss excitation is transferred even
without such a precise matching, due in part to the ability of the
two-level systems on the receiver side to exchange energy with the
oscillator.
Even though we sought to minimize energy exchange with the oscillator
in the numerical calculation, one can see a relatively small modulation
of the oscillator probability distribution in association with the
excitation transfer (Figure \ref{Trans1a}).


\epsfxsize = 3.50in
\epsfysize = 3.00in
\begin{figure} [t]
\begin{center}
\mbox{\epsfbox{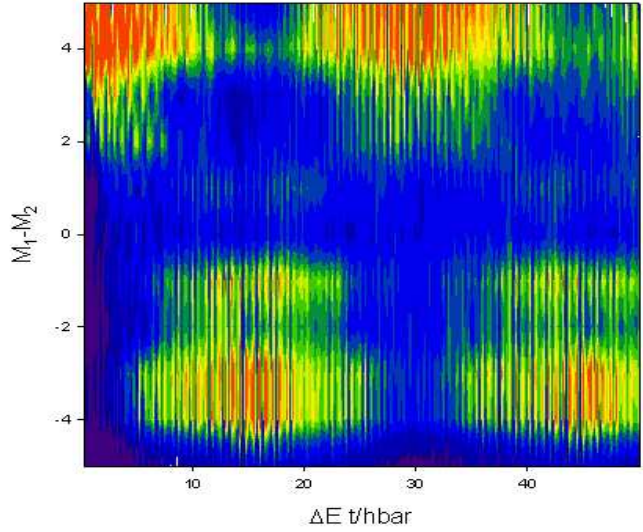}}
\caption{Demonstration of excitation transfer between the first set of two-level systems and
the second set of two-level systems in the case of infinitely fast loss.
The calculation was performed with two sets of 5 two-level systems, where
the undressed energy of the receiver-side system is $\Delta E_2 = 1.01 \Delta E_1$.
The coupling on the donor side is $V_1e^{-G} \sqrt{n_0}/\Delta E=0.2$, and the 
coupling on the receiver side is $V_2 \sqrt{n_0}/\Delta E = 0.6$. 
The oscillator energy was taken to be $\hbar \omega_0 = \Delta E_1/27$, and we
assumed an excitation level $n_0 = 10^8$.
In this calculation, the full lossy model Hamiltonian was diagonalized, and a
dynamic solution was developed from the eigenstates which shows the excitation
transfer effect with relatively small energy exchange with the oscillator.
Shown is the logarithm of the probability distribution for $M_1-M_2$ as a function of time.
One observes a fast excitation transfer effect in this case, with most of the
excitation being transferred from the first set of two-level systems to the
second.
The probability distribution associated with the oscillator is shown in the following figure.
}
\label{Trans1}
\end{center}
\end{figure}


\epsfxsize = 3.50in
\epsfysize = 3.00in
\begin{figure} [t]
\begin{center}
\mbox{\epsfbox{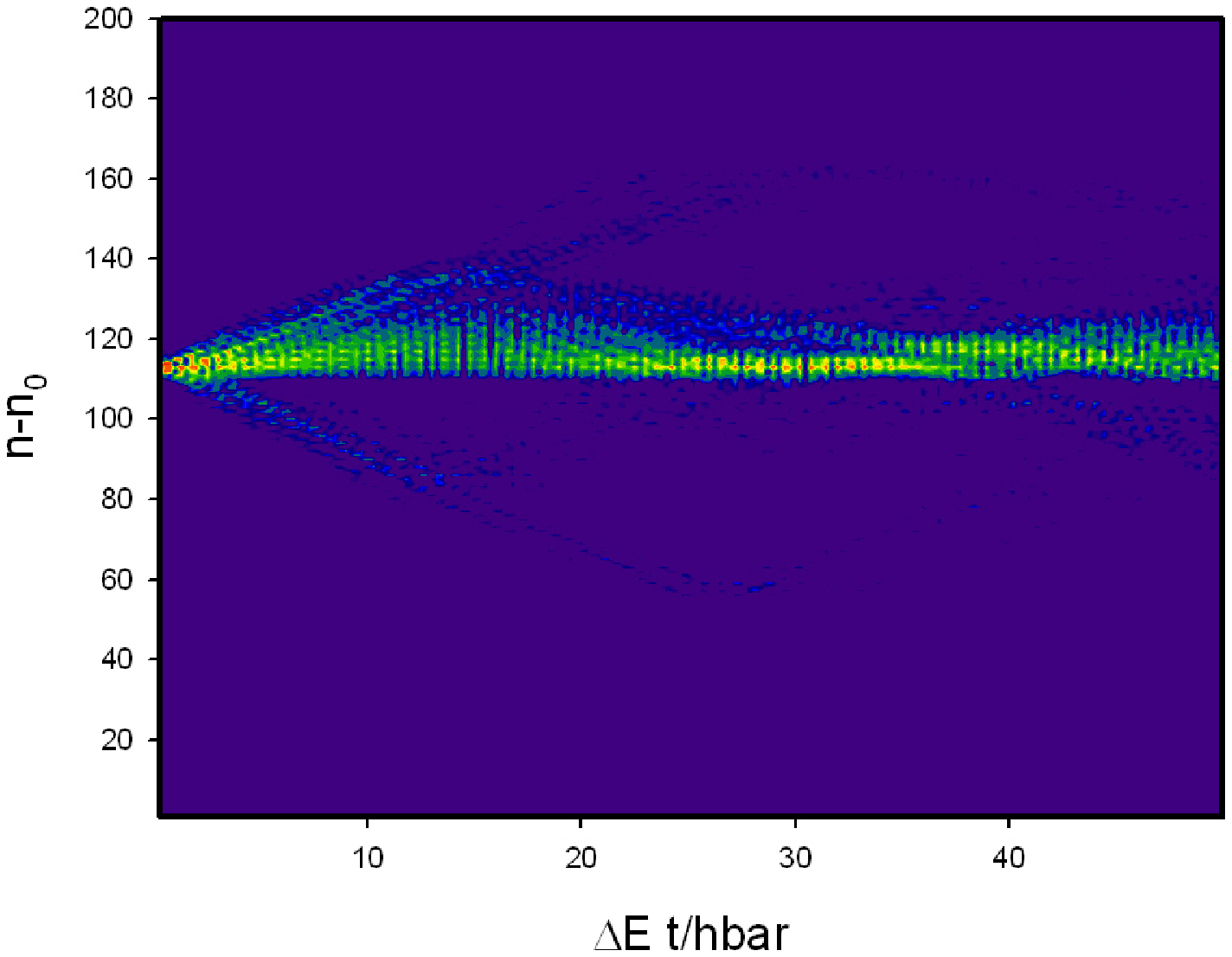}}
\caption{Logarithm of the probability distribution of the oscillator states as a function of time for the calculation
described in the previous figure.  
One sees that there is a small exchange of oscillator energy in association with the excitation
transfer effect. }
\label{Trans1a}
\end{center}
\end{figure}


\subsection{Excitation transfer: Strong coupling}

  We consider excitation transfer in the case of strong coupling on the
receiver side, and weak coupling on the donor side, in Appendix D.  
As before, we adopt an approximate solution which depends on the oscillator number $n$
according to $e^{i n \theta}$.
In addition, we assume a sinusoidal solution $e^{i (M_2-M_1)\phi}$
which depends on the difference between the level of excitation of
the two sets of two-level systems.
An oscillatory solution of the form $e^{i (M_2-M_1)\phi}$ becomes exact in
model in which the coupling terms in the Hamiltonian are taken to be
independent of $n$, $M_1$, and $M_2$, which makes the model periodic
in $(n,M_1,M_2)$-space as discussed in Section VI.
The variable $\phi$ is conjugate to $M_2-M_1$, and can be identified with
the momentum for the coupled system propagating in the $(M_2-M_1)$-direction.
In this case, the self-energy is given approximately by

\begin{equation}
E(\theta,\phi)
~\approx~
E_0 + {\Delta E \over 2}
-
2 \sqrt{n_0} \cos(\theta)
\left [
V_2  \sqrt{S_2^2 - (M_2)_0^2}
 +
V_1 e^{-G} \sqrt{S_1^2-(M_1)_0^2}
\cos(2 \phi)
\right ]
\label{trnsstrng}
\end{equation}

\noindent
where

$$
E_0 
~=~ 
\Delta E_1 (M_1)_0 + \Delta E_2 (M_2)_0 + \hbar \omega_0 n_0
$$

\noindent
and where $\Delta E_1 = \Delta E_2$ appears here as $\Delta E$.
The self-energy term from this model is the same as we obtained in the 
previous section.  A new term appears here that is associated with
indirect coupling between the two sets of two-level systems.
The excitation transfer rate (group velocity) is

\begin{equation}
{d \over dt} (M_2-M_1)
~\sim~
{1 \over \hbar} {dE \over d \phi}
~=~
{4 V_1 e^{-G} \sqrt{n_0} \cos(\theta) \sin (2 \phi) \sqrt{S_1^2 - (M_1)_0^2} \over \hbar}
\end{equation}

\noindent
Excitation transfer in this case proceeds freely with a maximum rate occuring
when $|\cos(\theta) \sin(2 \phi)|=1$.
%

\newpage
\section{Approximate periodic model and results}

We have considered a variety of models including loss at the two-level
system energy, and these models exhibit enhanced rates for excitation transfer
and energy exchange.
In the limit that the oscillator excitation is large (so that $n$ is large), and
the number of excited states of the two-level systems is also large, it is possible
to develop a different kind of model which takes advantage of the approximate
periodicity of the system in $(n,M_1,M_2)$-space.
The development of periodic solutions in the large $n$ limit has been studied in the closely related
spin boson model by Cibils et al.\cite{Cibils1991}
Calculations of the associated energy band structure leads to results which
support the conclusions obtained in the previous sections.

\subsection{Approximate periodic model}

In the models under discussion, there are three degrees of freedom corresponding
to the oscillator excitation ($n$), and the degree of excitation of the two-level 
systems ($M_1,M_2$).
Energy increases with increasing oscillator number and Dicke number, so that the
system cannot be periodic in the direction of increasing energy.
However, there are two other directions in which the Hamiltonian is approximately
invariant with respect to appropriate translations in $n$, $M_1$, and $M_2$.
If we work with an approximate Hamiltonian with interaction terms that are locally invariant,
then this new approximate model is periodic.
For example, such an approximate Hamiltonian can be written as

{\small

\begin{equation}
\hat{H}
~=~
\Delta E_1  M_1
+
\Delta E_2  M_2
+
\hbar \omega_0 n
-
{i \hbar \over 2}\hat{\Gamma}(E)
+
\Delta E_1
g_1
(\hat{\delta}_+^{(1)} + \hat{\delta}_-^{(1)})
(\hat{\delta}_+       + \hat{\delta}_-)
+
\Delta E_1
g_2
(\hat{\delta}_+^{(2)} + \hat{\delta}_-^{(2)})
(\hat{\delta}_+ + \hat{\delta}_-)
\end{equation}
}

\noindent
Here, we have defined two dimensionless coupling constants according to

\begin{equation}
g_1 
~=~ 
{V_1 e^{-G} \sqrt{n}_0
\sqrt{S_1^2-(M_1)_0^2}
\over 
\Delta E_1}
\ \ \ \ \ \ \ \ \ \ \ \ \ \
g_2 
~=~ 
{V_2 \sqrt{n}_0
\sqrt{S_2^2-(M_2)_0^2}
\over 
\Delta E_1}
\end{equation}

\noindent
To correspond with the models under discussion above, the reduced coupling
constant $g_1$ will be small, and the reduced coupling constant $g_2$ will be
large. 
The various $\hat{\delta}$ functions that appear are reduced raising and lowering
operators which act according to

\begin{equation}
\hat{\delta}_\pm
|n\rangle
~=~
|n \pm 1 \rangle
\ \ \ \ \ \ \ \ \
\hat{\delta}_\pm^{(1)}
|S_1,M_1\rangle
~=~
|S_1,M_1 \pm 1 \rangle
\ \ \ \ \ \ \ \ \
\hat{\delta}_\pm^{(2)}
|S_1,M_2\rangle
~=~
|S_1,M_2 \pm 1 \rangle
\end{equation}

\noindent
This Hamiltonian is invariant under translations of the form

\begin{equation}
\hat{T}_1 ~=~ \hat{\delta}^{(1)}_- + {\Delta E_1 \over \hbar \omega_0} \hat{\delta}_+
\ \ \ \ \ \ \ \ \ \
\hat{T}_2 ~=~ \hat{\delta}^{(2)}_- + {\Delta E_2 \over \hbar \omega_0} \hat{\delta}_+
\end{equation}

\noindent
as long as  
$\Delta E_1/\hbar \omega_0$
and 
$\Delta E_2/\hbar \omega_0$
are integers, and also as long as the loss operator is consistent
with periodicity.

\subsection{Numerical result}

We have used this model to investigate excitation transfer both on, and off, resonance.
We would like to confirm the approximate results for excitation transfer that we found
in the strong coupling limit in the previous section.
In addition, we find that the model predicts fast excitation transfer rates
when the energies of the two-level systems are not precisely matched, which is
interesting.
For the computations discussed here, we have adopted the approximate wavefunction
discussed in Appendix B.

  We consider a set of calculations done with the periodic model in which the
bare receiver-side transition energy is matched to a specific integer number of oscillator
quanta

$$\Delta E_2 = 101 \, \hbar \omega_0$$

\noindent
and that different combinations of donor-side transition energy and oscillator
quanta are matched to the receiver-side transition energy

$$\Delta E_1 + \Delta n \hbar \omega_0 ~=~ \Delta E_2$$

\noindent
We assume that the two dimensionless coupling strengths are

$$
g_1 = 0.01 
\ \ \ \ \ \ \ \ \ \ \ \ \ \ \ 
g_2 = 10.0
$$

\epsfxsize = 4.00in
\epsfysize = 3.00in
\begin{figure} [t]
\begin{center}
\mbox{\epsfbox{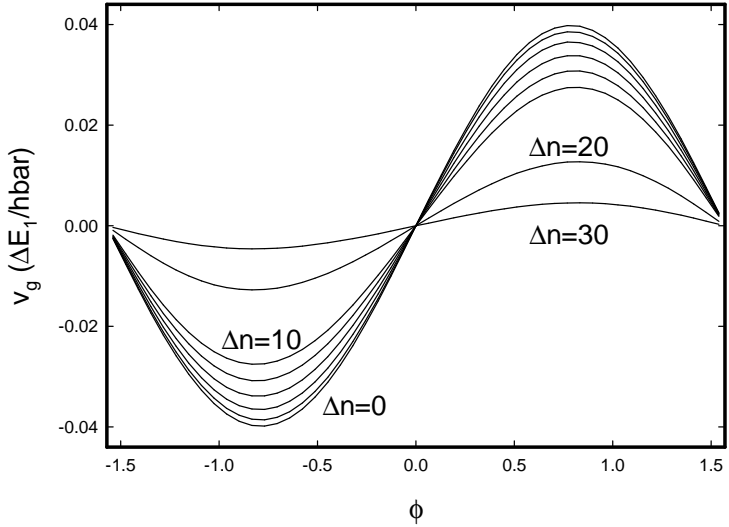}}
\caption{Group velocity calculated for a set of models based on the approximate periodic Hamiltonian
discussed in Section VI.  In these models, the receiver-side two-level system energy is
matched according to $\Delta E_2 = 101 \hbar \omega_0$, and the donor side is low by an
even number of oscillator quanta $\Delta n$ (0, 2, 4, 6, 8, 10, 20, and 30).  We have assumed
a restricted set of basis states consistent with the approximation discussed in Appendix B,
consistent with a loss model that increases rapidly with increasing available energy.
We have selected the lowest energy band, corresponding to $\theta \approx 0$, which would
be the band with the lowest loss in the case of the dipole coupling model described in
Appendix A.  Results are shown for the group velocity as a function of $\phi$, using
the definition of $\phi$ used in Section 5 and Appendix D.}
\label{vgroup2}
\end{center}
\end{figure}

\noindent
In a specific calculation, we obtain a moderate number of energy bands
which can be understood as originating from an approximate local $e^{in\theta}$
dependence which must match periodic boundary conditions.
Within this scheme, one of the bands will have the lowest loss (such as
near $\theta=0$ as discussed briefly in Appendix C).
Most interesting is the excitation transfer rate computed for the band
that corresponds most closely to the $\theta=0$ solution.
The lowest energy band corresponds most closely to this solution, and the
associated self-energy is close to what we would estimate using the results
from the local approximate solution given in the previous sections [Equations
(\ref{enexch}) and (\ref{trnsstrng})].
We present results for the excitation transfer rate from a set of calculations
shown in Figure \ref{vgroup2}.
One sees that the transfer rate on-resonance is very similar to the excitation
transfer rate off of resonance.
In this case, we see that the energy mismatch between $\Delta E_1$ and
$\Delta E_2$ must be about 14\% in order for the excitation transfer rate to drop
by a factor of 2 from its maximum value.

\subsection{Comparison with approximate result}

  We may check the approximate result that we obtained in Appendix D for the maximum
excitation transfer rate on resonance against the numerical result presented here. 
The energy estimate that we obtained in Equation (\ref{trnsstrng}) 
can be written in terms of the dimensionless
coupling constants as

\begin{equation}
E(\theta,\phi) 
~=~ 
E_0 + {\Delta E \over 2} 
- 2 g_2 \cos(\theta) \Delta E_1
- 2 g_1 \cos(\theta) \cos(2 \phi) \Delta E_1
\end{equation}

\noindent
The approximate result for the excitation transfer rate can then be written 
as

\begin{equation}
{d \over dt} (M_2-M_1)
~\sim~
 4 g_1 \cos (\theta) \sin (2 \phi)
{\Delta E_1 \over \hbar}
\end{equation}

\noindent
The numerical solution corresponds to $\cos(\theta)=1$, and we have used $g=0.01$. 
Consequently, the approximate solution evaluates to

\begin{equation}
{d \over dt} (M_2-M_1)
~\sim~
0.04 \, \sin (2 \phi)
{\Delta E_1 \over \hbar}
\end{equation}

\noindent
This corresponds well with the $\Delta n = 0$ result plotted in Figure \ref{vgroup2}.

The reason that the approximate result works so well in this comparison is that
the numerical result corresponds to the periodic model with small (but finite)
$\hbar \omega_0$ while the approximate result corresponds to zero 
$\hbar \omega_0$.


\newpage
\section{Connection with experiment}

   The models presented here were developed in response to experiments
that appear to show an excess heat effect in metal deuterides.
A key feature of these experiments of interest to us is that they 
seem to present examples of nuclear reactions in which the reaction
energy is not expressed as energetic particles, as would be expected
in vacuum reactions.
In one application of the basic model, 
the upper and lower levels on the donor side may be molecular D$_2$ and $^4$He states.
If the excitation energy of D$_2$ relative to $^4$He is transferred elsewhere, such
as to highly excited states in Pd, then reaction energy is not expressed in energetic
p+t or n+$^3$He channels as occurs in vacuum.
Rapid excitation transfer from one site to another in the receiver nuclei would allow
the energy to be exchanged coherently with the highly excited phonon mode, 
or incoherently with other phonon modes, a small number of quanta at a time.

The number of issues associated with such a proposal is quite large, and we
do not have the opportunity here to address each of them.
We choose to focus here on phonon exchange, and screening as it contributes to 
estimates of the reaction rate.

%

\subsection{Phonon exchange}

  The inclusion of phonon exchange a nuclear reaction matrix element has been discussed 
previously by Chaudhary and Hagelstein.\cite{ChaudHag2006}
From this work, one may think about phonon exchange from the point of view of the 
lattice simply.
When two deuterons tunnel sufficiently close for a strong force interaction to occur, together 
they appear the rest of the lattice as a localized mass 4 and charge 2 object.
A direct transition to $^4$He would not be expected to produce significant phonon exchange,
since helium also appears to the rest of the lattice as a localized mass 4 and charge 2 object.
For phonon exchange to occur, there needs to be a change in the local charge (and hence force
constants) or mass.
  Consequently, we consider a second-order transition in which 

$${\rm D}_2 ~\leftrightarrow~ {\rm n}+^3{\rm He} ~\leftrightarrow~ ^4{\rm He}$$  

\noindent
In such a second-order transition, we need to sum over contributions from all possible n+$^3$He intermediate
states

\begin{equation}
V_1(\epsilon) e^{-G} \sqrt{n_0}
~\rightarrow~
\sum_j
{
\langle \Phi  [^4{\rm He}]         | \hat{V}_n | \Phi_j[^3{\rm He}+{\rm n}] \rangle
\langle \Phi_j[^3{\rm He}+{\rm n}] | \hat{V}_n | \Phi[{\rm D}_2]            \rangle
\over
(\epsilon - E_j)
}
\label{indirect}
\end{equation}

\noindent
where $\hat{V}_n$ is the strong force interaction operator.
Here individual nuclear matrix elements calculated including phonon exchange effects as
outlined in Chaudhary and Hagelstein.\cite{ChaudHag2006}

  We would expect to encounter destructive interference in the summation over intermediate n+$^3$He
states for off-resonant states, which is equivalent to a localization of the neutron in the vicinity
of $^3$He, and we would not expect to see significant phonon exchange as the virtual n+$^3$He state
would also look like a mass 4 charge 2 object to the rest of the lattice.
We would expect significant phonon exchange only for n+$^3$He states in which the relative energy
is on the order of the vibrational energy of the $^3$He nucleus associated with the highly excited
phonon mode.
A schematic (exaggerated) of the corresponding classical trajectories in such a case are illustrated
in Figure \ref{traj1}.
An analogous deviation of trajectories would be expected for intermediate p+t states; however,
we have not included them since at such low relative energies one would expect a tunneling
hindrance due to the Coulomb barrier.

\epsfxsize = 4.00in
\epsfysize = 3.00in
\begin{figure} [t]
\begin{center}
\mbox{\epsfbox{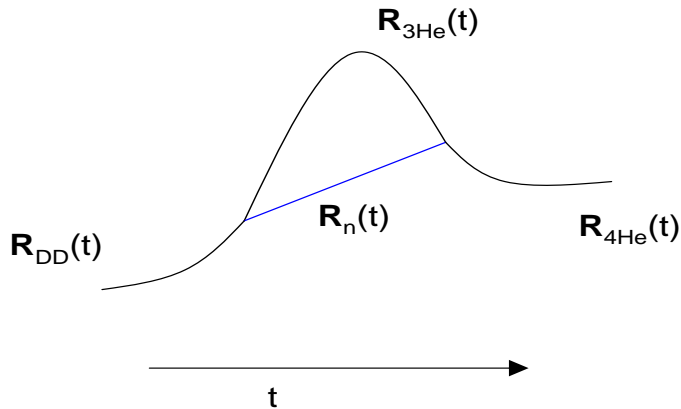}}
\caption{Schematic of classical trajectories corresponding to intermediate states in which
the relative neutron and $^3$He energies are on the order of the $^3$He oscillation energy
due to a highly excited phonon mode.  Prior to the first interaction, two deuterons are
initially localized with a center of mass location as indicated.  After the first interaction,
the $^3$He nucleus sees essentially the same forces but responds with a lighter mass, while
the neutron follows a straight line trajectory.  Significant contribution to the second-order
interaction will only come from such states in which the classical trajectories intersect 
later on so that there is a likelihood of a second interaction.  After the second interaction,
the final state $^4$He nucleus follows a trajectory different that followed by the initial
localized D$_2$ center of mass trajectory due to the interaction.  The difference in trajectories
due to the interactions and separation of the neutron and $^3$He in the intermediate state
corresponds in the classical picture to phonon exchange having occured.
}
\label{traj1}
\end{center}
\end{figure}

We might expect that the situation on the receiver side should be analogous.  
Hence, we propose that an initial ground state receiver nucleus 
makes a phonon-induced indirect transition
to a near-resonant excited state  
by going through intermediate daughter plus neutral states.
Only intermediate states of this kind with relative kinetic energy on the order
of the local initial vibrational energy would be expected to be able to exchange
phonons.


%
%

\subsection{Excitation transfer rate and consistency with experiment}

Of interest is whether the excitation transfer model can help explain the
excess heat effect discussed in the Introduction.
To assess this, we face several issues.
We require excitation to occur at a rate fast enough so that the associated
excitation energy transferred per unit time is predictive of the excess power.
We also require that the energy exchange between the excited nuclei on
the receiver side occurs fast enough to convert the excitation once transferred,
but not so fast that the receiver-side is no longer in the strong coupling regime.
In addition, we need an estimate of the hindered coupling strength on the 
donor side.
This is made difficult due to the present situation in which experiment shows
clearly screening effects which are not at present matched by theory.

  Nevertheless, it is reasonable to embark on the discussion here, in spite of these
difficulties.  
We begin with the ansatz that the energy exchange effect is fast enough that excitation
transfer is rate limiting, but slow enough that the system remains in the strong coupling
limit.
The maximum rate per donor-side two-level system
at which excitation transfer can occur in the case of a single mode and
single receiver-side excited state in the model is

\begin{equation}
\gamma_{max}
~=~
{
\left [
\displaystyle{1 \over 2} \displaystyle{d \over dt} (M_2-M_1)
\right ]_{max}
\over
2 S_1
}
~=~
{
\left [
\displaystyle{2 V_1 e^{-G} \sqrt{n_0} \sqrt{S_1^2-M_1^2}  \over \hbar}
\right ]_{max}
\over
2S_1
}
~=~
{V_1 e^{-G} \sqrt{n_0}  \over \hbar}
\end{equation}

\noindent
It seems a reasonable conjecture to assume that concurrent excitation transfer 
to other excited states should be possible, producing
a larger rate for excitation transfer from the excited donor state.
It may be the case as well the donor levels could concurrently transfer excitation
in association with other highly excited modes.
Such effects would further increase the total rate for excitation transfer over
the single-mode and single receiver-side excited state estimate.
The maximum power that can be associated with the basic single-mode, single receiver-side
excited state case is

\begin{equation}
P_{max}
~=~
N_{DD}
\Delta E_1
{V_1 e^{-G} \sqrt{n_0}  \over \hbar}
\end{equation}

\noindent
where $N_{DD}$ is the number of molecular D$_2$ states participating.  We note that our
simple estimate below will be referenced to the ground state, where in a more accurate
calculation we should take into account that other rotational and vibrational
states will be occupied as well.

   Since the tunneling factors are very small, it has proven difficult for any theory to
successfully account for the rates associated with the excess heat effect.
Consequently, it may be illuminating here to examine a representative example under the
assumption that only a single mode and single receiver-side excited state is involved.
A variety of experimental and theoretical results leads to the conclusion that it is the 
near-surface region as being active in a successful Fleischmann-Pons experiment.
Calculations on bulk palladium hydride\cite{Sun89,Liu89,Lam89,Chris89,Wang89,Wei90,Swit91} 
indicate that the mean separation between
protons is substantially greater than for molecular H$_2$, so that one would not expect 
deuterium to be present in states resembling molecular D$_2$ in the bulk.
A simple way to think about this is that molecular D$_2$ anti-bonding states are occupied 
in palladium deuteride, which results in the absence of a potential minimum below
1 \AA \ separation.\cite{Wei90}
Hence, we need to look elsewhere for a suitable host environment, which leads us to the
cathode surface where one can typically find dendritic structures produced by codeposition.
Codeposition of Pd at high hydrogen or deuterium chemical potential produces a high defect
concentration since single site defects are stabilized,\cite{Fukai} 
and we conjecture that D$_2$ in states should occur under conditions where antibonding
states do not have such high occupation as in the bulk.
We assume for this discussion that the outer 10 microns of the cathode consists of 
codeposited material with a high defect concentration 
in which molecular D$_2$ is present in relatively high concentrations (10$^{-4}$), 
leading to a population $N_{DD}$ on the order of $4 \times 10^{15}$ D$_2$ molecules
for a 1 cm$^{2}$ cathode surface area.

  We require next a parameterization of the interaction matrix element [realistic
calculations based on Equation (\ref{indirect}) are in progress].  
Two deuterons must tunnel together to interact, which suggests that the interaction
matrix element should include a Gamow factor and volume factors according to

\begin{equation}
V_1 \sqrt{n_0} e^{-G}
~\rightarrow~ 
\sqrt{v_{nuc} \over v_{mol}} 
~e^{-G(U_e)}
~u_{DD}
\end{equation}

\noindent
where $v_{mol}$ is the relevant volume associated with molecular D$_2$, and where $v_{nuc}$ is
the relevant volume associated with the nucleon distribution when strong force interactions
occur (we use $\sqrt{v_{nuc}/v_{mol}}$ to be $8.5 \times 10^{-8}$ here).  
We have written the tunneling factor with an explicit dependence on the screening potential $U_e$
in recognition that our result will depend on what screening model is used.
The parameter $u_{DD}$ is the residual second-order interaction between two deuterons leading
to phonon exchange and $^4$He formation.
We adopt a provisional value of 1 MeV for $u_{DD}$.

The tunneling factor in the case of molecular D$_2$ can be estimated from
fusion rate calculations presented in the literature\cite{Koonin,Shimamura} to be
about $1.7 \times 10^{-36}$.
Experiments by the LUNA collaboration on low energy deuteron-deuteron fusion reactions show enhancements in the
yield which can be fit through the use of a screening potential $U_e$.\cite{Raiola2004,Rolfs2004,Raiola2005}
For insulators the screening potential values found are low ($<30$ eV), which is 
similar to the result for molecular D$_2$.
Screening energies for metals are in the range of 130 eV to 800 eV.
The experimental results appears to follow an empirical scaling law based on a Debye screening
formula.
Czerski et al\cite{Czerski} have recently published a calculation of the fusion rate per molecule as
a function of the screening potential, from which one sees a roughly 30 order of magnitude increase
at 100 eV, and 40 order of magnitude increase at 200 eV.

  In light of this, we do not have a good estimate for the Gamow factor appropriate to the outer cathode
surface region, which makes the task of predicting a reaction rate difficult at this time. 
However, we can develop a consistency check to see whether the screening parameter required by
the model to match the excess heat is consistent with low energy fusion experiments.
The largest screening parameter is required under the assumption of a singly highly excited
phonon mode, and single receiver-side excited state.
We match the excitation transfer rate to a representative excess power, and solve for the screening
parameter $U_e$

\begin{equation}
N_{DD}
\Delta E_1
\sqrt{v_{nuc} \over v_{mol}} 
{u_{DD} \over \hbar}
~e^{-G(U_e)}
~=~
0.3~{\rm watt}
\end{equation}

\noindent
Solution of this equation produces a (near zero relative energy) screening energy $U_e$ of about 115 eV, 
which is below the screening energies obtained from (keV relative energy) deuteron-deuteron fusion
experiments in metals.
According to the modeling of Czerski et al\cite{Czerski}, one would expect the screening energy to
depend on the relative deuteron energy.  
In a model calculation relevant to PdD, the screening energy at low (1 eV) relative energy ($U_e \sim$ 55 eV)
is about 60\% of the screening energy at high (10 keV) relative energy ($U_e \sim$ 90 eV). 
Experiment for this case is consistent with a screening energy of 800$\pm$90 eV. 
At present, no model can account for the large screening energies observed experimentally.
However, the (near zero relative energy) screening energy $U_e$ required here for consistency seems quite plausible.

\subsection{Energy exchange with the lattice}

   In the model that we have proposed, the initial (slow) excitation transfer is followed
by subsequent (fast) excitation transfer reactions among the receiver-side nuclei.  
With each individual receiver-side excitation transfer reaction there is the possibility
of coherent and incoherent phonon exchange with the lattice.
We do not have an estimate for the rate of incoherent exchange from this model, but
we can compare the coherent rate required with our strong coupling result to assess whether 
the associated coherent rate is sufficient.
To proceed, we form the ratio $r$ of the power associated with excitation transfer and 
the power associated with energy exchange  

\begin{equation}
r 
~=~ 
{  
\left [
\displaystyle{\Delta E_1 \over 2} \displaystyle{d (M_2-M_1)\over dt}
\right ]
\over
\left [
\hbar \omega_0 \displaystyle{d n \over dt}
\right ]
}
~=~
{\Delta E_1 \over \hbar \omega_0}
{V_1 \over V_2}
{\sqrt{S_1^2 - M_1^2} \over \sqrt{S_2^2 - M_2^2}}
\cot(\theta) \sin (2 \phi)
~
e^{-G}
\end{equation}

\noindent
In the event that the two powers are matched at the maximum excitation transfer rate [$\sin(2 \phi)=1$], 
then $r=1$, and we may write

\begin{equation}
\tan (\theta)
~=~
{\Delta E_1 \over \hbar \omega_0}
{V_1 \over V_2}
{\sqrt{S_1^2 - M_1^2} \over \sqrt{S_2^2 - M_2^2}}
~
e^{-G}
\end{equation}

\noindent
There is no information available from experiment at present 
as to what phonon mode frequencies are involved.
For phonon mode frequencies in the kHz to MHz regime, the combination of
the Gamow factor, and volume factor in $V_1$, lead to solutions
where $\theta$ is very small under most conditions.
We conclude from this that in the strong coupling limit the excitation
energy can readily be transferred to the oscillator, 
and that this argument supports the ansatz of the previous subsection.

\newpage
\section{Summary and conclusions}

%

  We considered basic models in which two sets of two-level systems are coupled to a common
highly excited oscillator,
where the two-level system transition energy is much greater than the oscillator energy.
We focused on the case of linear coupling with the oscillator (qualitatively similar results can
be obtained with nonlinear coupling).  
These models exhibit an excitation transfer effect, 
in which excitation from one set of two-level systems is transferred to the other set.
This effect is a consequence of indirect coupling between the two sets of two-level systems.
We have demonstrated the existence of this excitation transfer effect 
through a numerical calculation, and also through approximate diagonalization.
We also demonstrated the existence of an energy exchange effect, in which energy from
the two-level systems is transferred coherently to the oscillator, even though the
energies may be incommensurate.
This effect comes about because excitation transfer among the two-level systems
occurs in association with oscillator exchange, and coherence can be maintained
over the course of many such transfer and exchange processes.
Both effects are weak in the lossless case, and require precise resonances between
the two-level systems (for excitation transfer), and two-level systems and multiples
of the oscillator energy (energy exchange).

We then discussed a version of the model augmented with loss effects at the two-level transition energy.
The inclusion of loss changes the problem qualitatively, due to the elimination of
destructive interference effects between different pathways involved in coupling between
basis states.
When loss is present, the rates for excitation transfer and energy
exchange are greatly increased (in comparison with the lossless case), and 
depend weakly on the specific loss model used.
Consequently, we have introduced an approximate restricted wavefunction that
omits states most strongly impacted by loss for a loss model in which decay rates
increase strongly with energy.
We take advantage of this approximation to develop numerical and analytic results
for the excitation transfer and energy exchange effects.
The analytic results developed in the strong coupling limit are in good agreement
with results obtained using a periodic model.


These models were developed in response to experiments in which excess heat is
observed at levels much greater than can be accounted for by chemical processes,
in which no energetic particles are observed.
Measurements of $^4$He correlated with the excess energy lead one to conclude that
some new kind of nuclear process is involved, and the reaction $Q$-value determined
from experiment is about 24 MeV consistent with 

$${\rm d} + {\rm d} ~\to~ ^4{\rm He} + 24 ~{\rm MeV}~ {\rm (heat)} $$

\noindent
This conjecture has been controversial since there have been no previous observations
of reactions that work this way in nuclear physics, and because one normally expects
the reaction energy to be expressed in terms of energetic particles.
However, if the local reaction energy of a deuteron-deuteron reaction can be transferred
elsewhere (such as described in the simple models discussed in this paper), then the
situation changes.
The idealized models described in this paper then correspond to new reaction mechanisms
that have not been seen previously in nuclear physics, and which would behave 
differently.

We examined the reaction rate for excitation transfer in connection with
excess heat in a Fleischmann-Pons experiment, and we conclude that theory and experiment
can be consistent as long as a strong screening effect is present.
Recent experimental work on low-energy deuteron-deuteron fusion reactions have
shown such a strong screening effect in metals (but not insulators) 
in the few keV range, and it seems reasonable to expect that this effect will
persist at lower energies.


\newpage
\section*{Appendix A: Loss}

   We are interested in this appendix in loss mechanisms available to the 
coupled quantum system 
(made up of two-level systems for nuclear states and an oscillator for a highly excited phonon mode) 
near the two-level transition energy and above.
We recognize two qualitatively different kinds of decay mechanisms: 
(1) decay of excited nuclear states associated with the two-level systems; and
(2) induced decay processes when the coupled system has allowed decay modes.
A presumption that we have made in putting forth a model based on two-level systems
is that the nuclear states associated with the levels of the two-level systems are
stable at least on a timescale of the excitation transfer dynamics.
In the case of molecular D$_2$, numerous calculations have appeared showing
slow decay rates (even with screening at the levels discussion in Section VII).\cite{Koonin,Czerski}
The issue of which specific states on the receiver side is outside of the scope of this
paper; for the purposes of the discussion we assume that suitable reasonably stable 
excited states exist and can be accessed.

   Consequently, our attention here is focused on decay mechanisms available to the coupled
quantum system.
In Section III, we discuss the use of perturbation theory to obtain an estimate for the
coupling associated with excitation transfer in which indirect coupling proceeds 
through off-resonant basis states with energy eigenvalues much less than the initial and final
basis state energy eigenvalues. 
One would expect such off-resonant states to be able to decay through whatever 
mechanisms are available at the energy defect (which is the transition energy of the
two-level systems, which in our discussion is on the MeV scale).
If the coupled system (oscillator and two-level systems) is in a strong coupling
regime, then a discussion in terms of basis states of the uncoupled system is not
helpful, and we need to think in terms of transitions to accessible lower energy
states of the coupled system.
In either case, we would expect the coupled system to dissipate a large energy
quantum. 
In this appendix we are interested in specific dissipation mechanisms and models. 
In the next section we will be concerned with the impact of loss on the wavefunctions.

\subsection*{Decay mechanisms}

   The major decay mechanisms that we might expect include:

\begin{itemize}

\item Induced nuclear decay, in which one or more energetic particles is emitted.

\item Induced electron recoil, in which a K-shell electron recoils from a nucleus.

\end{itemize}

\noindent
One might conjecture that there should exist another loss mechanism in which a 
large energy quantum is dissipated through a large number of low energy decays,
such as electron promotion, phonon generation, atom ejection, and so forth.  
In our view, if such a decay path existed, evidence for it should have been seen
in association with inner-shell fluorescence experiments and experiments involving
conventional nuclear decay.
One can make theoretical arguments as to why this kind of effect is unlikely, and
so we will focus our discussion on low-order energetic decay mechanisms.

  If one unit of transition energy is available, 
we might expect to lose excited states of the two-level systems to the above decay mechanisms.
If the transition energy is great enough (keeping in mind that we may be interested in
schemes based on HD where the transition energy is about 5.5 MeV), we may lose lower states as well.
Upon reflection, one might also expect to see decays of nuclei that make up the host lattice, 
since one would expect phonon exchange to occur with the highly excited oscillator in
association with an induced disintegration.
The reason that we are concerned with such issues is that the specific loss mechanisms end up impacting the
wavefunction, as we discuss in Appendix B.
In this regard, loss mechanisms for donor or receiver nuclei have similar consequences for
the wavefunction of the coupled system.
It is then convenient here to consider the donor and receiver nuclei as being minor constituents
in a host lattice, and focusing on (energetic) loss mechanisms induced in the host lattice through
phonon exchange.

   In the introduction, we discussed excess heat experiments as producing excess heat but
not producing energetic charged particles. 
Yet here we are focused on loss mechanisms that involve the production of energetic
charged particles, which may seem counter-intuitive.
The resolution of this is that in our view the wavefunction seeks to avoid regions of high loss 
(as discussed further in Appendix B), 
and by doing so achieves in these models stronger coupling, 
and faster excitation transfer and energy exchange rates.
However, we note that there are many experiments with metal deuterides in which excess heat is not observed, 
in which a variety of energetic particles are present, 
and where some of the loss mechanisms under discussion seem to be relevant.  
A systematic understanding of these loss mechanisms in our view is a prerequisite for
a comprehensive understanding of the anomalies.

\subsection*{Phonon exchange}

The possibility that the coupled oscillator and two-level system may induce
disintegrations or electron recoil in nuclei that make up the host lattice
motivates us to consider the issue of phonon exchange and coupling
in more detail.  
%
%
We are familiar with the lack of phonon exchange in the M\"ossbauer effect, as
well as the breakdown of the M\"ossbauer effect when the recoil is strong.  
Phonon exchange through recoil in the presence of a highly excited phonon mode
has been studied.\cite{Gupta1974}
In the event that the nucleus disintegrates, there should also occur phonon exchange
with a highly excited phonon mode.
In support of this we consider a lattice with a highly excited phonon mode in which
a nuclear disintegration occurs under conditions in which the sudden approximation
is valid. 
In the initial state, the nucleus constitutes part of the lattice, and in the final
state the nucleus has broken up and the fragments are ejected.
In this case, the lattice has changed, and phonon exchange occurs in association
with this lattice change.
Phonon exchange in association with a force constant change 
is sometimes referred in terms of a Duschinsky mechanism,\cite{Duschinsky,Sharp,Faulkner}  
but the matrix formulation applies on equal footing to a mass change.
In the present example, nuclear disintegration in the sudden approximation results
in both a local mass change (to zero mass since the nucleus is no longer present)
as well as a change in force constants (where there used to be forces, none are
present in the final state).
Consequently, we consider phonon exchange in the case of induced nuclear disintegration
to occur through a Duschinsky mechanism.

   Induced alpha decay and fission processes would be expected to behave as a generic
disintegration following the arguments above.  
Induced beta decay processes will involve recoil and force constant change, and fit within the
scheme under consideration.
Induced K-shell electron recoil (in the non-M\"ossbauer limit) can produce phonon exchange
through the associated nuclear recoil if the electron energy is low, or through atomic displacement
if the electron is energetic.
In all cases, we would expect the possibility of phonon exchange with a highly excited
phonon mode.

\subsection*{Interaction Hamiltonian}

   As an example, we consider the situation in the case of induced
alpha decay and other decay processes involving nuclear fragments.
In this case, the development of an interaction Hamiltonian is related to 
(but simpler than) the approach presented recently by Chaudhary and 
Hagelstein,\cite{ChaudHag2006} since all that happens to the lattice is that a nucleus is 
lost at one site.
In general such a loss will produce coupling to essentially all of the
phonon modes, but we are only interested here in whatever coupling
occurs with the highly excited mode.
We would expect to end up with a single site interaction Hamiltonian of the 
general form (assuming that the decaying nucleus is not part
of either the donor or receiver two-level systems)

\begin{equation}
\hat{U}
~=~
\sum_{n,n^\prime}
\sum_{j,k}
|\phi_{n^\prime} \rangle
|\psi_k \rangle
U^{kj}_{n^\prime n}
\langle \psi_j |
\langle \phi_n |
\end{equation}

\noindent
The nuclear states here are denoted by $|\psi \rangle$, and the subscripts
$j,k$ range over all relevant initial and final nuclear states.
The oscillator states here are denoted by $|\phi \rangle$, and phonon exchange
occurs when $n \ne n^\prime$.

  We note that there should exist coupling matrix elements to essentially all 
available final states in general, but that energy conservation determines which
final state channels are open.
This is significant in the present discussion in that we should expect a much
different dissipation rate and set of products if the available energy is
``low'' (for example, 1 MeV) as compared to if it is ``high'' (for example, 30 MeV).
The induced disintegration rate in this regime will increase rapidly with increasing
energy as more decay channels open.
One would also expect there to be statistical factors present when the energy
becomes high enough so that two or more fast decays at different sites is
allowed.
We draw attention to this here as later on this rapid increase in decay with
energy will impact the form of the wavefunction we would expect.

\subsection*{Second-order interactions}

The most straightforward way to include this kind of decay into the model
Hamiltonian under discussion is through infinite-order Brillouin-Wigner
theory, in which one sector is identified in which the nucleus is in the ground state
and another sector is identified in which the daughter and fragments are present.
In this case, the associated second-order interaction is of the form

\begin{equation}
\hat{W} 
~=~
\hat{U}
[E - \hat{H}_0]^{-1} 
\hat{U}
\end{equation}

\noindent
An interaction appropriate to the idealized (and reduced) model can then be developed by
taking the expectation value over the ground state nucleus

\begin{equation}
\langle \psi_i | \hat{W} | \psi_i \rangle
~=~
\langle \psi_i |
\hat{U}
[E - \hat{H}_0]^{-1} 
\hat{U}
| \psi_i \rangle
\end{equation}

\noindent
Ultimately, we will end up with an interaction of the form

\begin{equation}
\langle \psi_i | \hat{W} | \psi_i \rangle
~=~
\sum_{n,n^{\prime \prime}}
|\phi_{n^{\prime \prime}} \rangle
W_{n^{\prime \prime}n}
\langle \phi_n |
\end{equation}

\noindent
This interaction will have a real part associated with the self-energy
as well as an imaginary part associated with loss if any loss channels
are open.
The augmentation of the idealized model to include loss
should contain terms of the form

\begin{equation}
-i {\hbar \over 2}
\hat{\Gamma}(E)
~=~
i \, {\rm Im}  \bigg \lbrace 
\sum_j \langle \psi_i | \hat{W} | \psi_i \rangle_j 
\bigg \rbrace
~=~
-i 
{\hbar \over 2}
\sum_{n,n^{\prime \prime}}
|\phi_{n^{\prime \prime}} \rangle
\Gamma_{n^{\prime \prime}n}
\langle \phi_n |
\end{equation}

\noindent
where the summation over $j$ includes all sites at which a disintegration might occur.

In the event that we worked with a model in which the Duschinsky mechanism resulted
in single phonon exchange, and the basic interaction $\hat{U}$ becomes proportional 
to $(\hat{a}+\hat{a}^\dagger)$, then we would expect the loss to depend on the
oscillator operators according to

\begin{equation}
-i {\hbar \over 2}
\hat{\Gamma}(E)
~\sim~
(\hat{a}+\hat{a}^\dagger)^2
\end{equation}


\newpage
\section*{Appendix B: Effect of loss on the states of the system}

In this appendix we are interested in the impact of loss on wavefunctions of
the coupled system, and in the development of approximate wavefunctions for
use in estimating rates for excitation transfer and energy exchange.
In the end, we propose an approximate wavefunction in which parts that are
most strongly impacted by loss are omitted.
The proposal for such an approximation is based on intuition that 
the probability amplitude arranges itself so as to avoid high-loss regions.
We begin by examining this in a simple lossy two-state system.
%
%
%

\subsection*{Two-state model with loss}

  We consider first the case of two generic quantum states, one lossy and one with no loss, that
are coupled.
We are interested in seeing what happens under conditions when the loss becomes large.
For such a problem, we may write

\begin{equation}
E
\left (
\begin{array} {c}
c_1 \cr
c_2 \cr
\end{array}
\right )
~=~
\left (
\begin{array} {cc}
H_1 & V \cr
V   & H_2 - i \displaystyle{\hbar \over 2} \gamma \cr
\end{array}
\right )
\left (
\begin{array} {c}
c_1 \cr
c_2 \cr
\end{array}
\right )
\end{equation}

\noindent
We can solve for $c_2$ in terms of $c_1$ to find

\begin{equation}
c_2 
~=~ 
\left [
{V \over E - H_2 + i \displaystyle{\hbar \over 2} \gamma}
\right ]
c_1
\end{equation}

\noindent
This can be used to develop a nonlinear eigenvalue equation

\begin{equation}
E 
~=~
H_1 
+
{V^2 \over 
E - H_2 + i \displaystyle{\hbar \over 2} \gamma}
\end{equation}

\noindent
There are two solutions in general.  In the event that the loss is
large, then the solutions divide up into a low-loss
solution and a high-loss solution.  For the low-loss solution, we
may write

\begin{equation}
c_2 
~\rightarrow~ 
-i {2V \over \hbar \gamma}
c_1
\ \ \ \ \ \ \ \ \ \ \ \ \ \ \
E 
~\rightarrow~ 
H_1
-i{2V^2 \over \hbar \gamma}
\end{equation}

\noindent
This is interesting since the probability amplitude for the low-loss solution
avoids the high-loss region.  The loss associated with the low-loss solution
scales as $1/\gamma$ in this limit.  Hence, the stronger the loss term, the
lower the loss of the coupled system.  This kind of behavior is reasonably
general, and provides us with some intuition about the more complicated 
coupled problem that we consider in what follows.

In the high-loss solution, the situation is reversed.  
In this case the probability amplitude seeks lossy regions, and avoids
low-loss regions. 
These solutions are less interesting to us since in general they decay rapidly,
and hence are unlikely to be relevant to the basic effects of interest in this
manuscript.

\subsection*{Weak coupling limit with strong loss}


\epsfxsize = 3.00in
\epsfysize = 2.25in
\begin{figure} [t]
\begin{center}
\mbox{\epsfbox{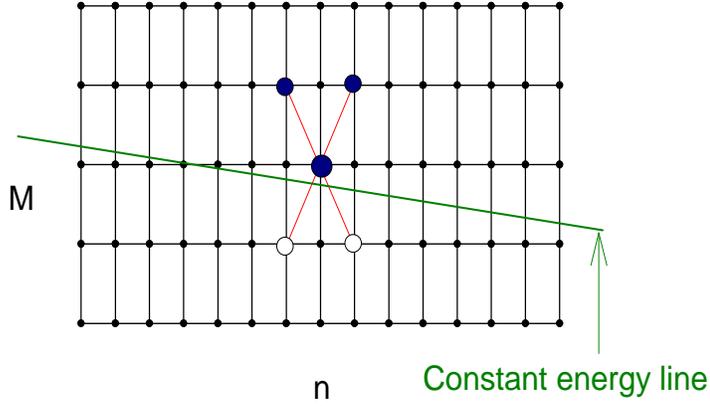}}
\caption{Schematic of wavefunction and constant energy line in the case of one set of two-level
systems coupled to an oscillator in the weak coupling limit.  
Illustrated are the different basis states which appear here as points in $(n,M)$ space.
In the weak coupling limit, the wavefunction is composed predominantly of a individual basis
state, which we denote here as a enlarged dark circle (that appears at the center of crossed lines).
Coupling occurs from this state to neighbors that differ by one unit of $n$ and $M$, which
is indicated by four lines originating from the enlarged dark circle, and circles indicating
probability amplitude.
A constant energy line is illustrated.
This constant energy line corresponds to constant $M \Delta E + n \hbar \omega_0$ plus a small
self-energy shift.
Coupling from the center point to the stable states above the constant energy line produce 
a small probability amplitude in weak coupling, which is indicated as smaller dark circles.
Coupling from the center point to the unstable states below the constant energy line produce 
a much smaller probability amplitude in weak coupling, which is indicated as smaller open circles.
}
\label{constE1}
\end{center}
\end{figure}


Consider next a model for the receiver side alone with loss, 
in the limit that the coupling between the two-level systems and oscillator are weak.
In this case, we could sensibly work with basis states in $(n,M)$ space that are discrete points, and
associate a loss with each point depending on the relevant energy eigenvalue $E$.
A schematic of this situation is illustrated in Figure \ref{constE1}.
The constant energy line in this figure is

\begin{equation}
E ~=~ \Delta E M + \hbar \omega_0 n + \Sigma(E)
\end{equation}

\noindent
where $\Sigma(E)$ is the self-energy, which is small in weak coupling.
The two lower basis states that are reached with linear coupling are below the
constant energy line, and we expect these to decay rapidly and hence have a small
probability amplitude.
Consequently, in the limit where the loss becomes infinite, we would expect
that in weak coupling the eigenfunction would be well approximated by the three
points above the constant energy line.

\subsection*{Intermediate and strong coupling}

  To extend our discussion to cases with stronger coupling, we need to rethink how we
approach the problem.  
The reason for this is that we have significant coupling present leading to
entanglement between many states, that entanglement is not lost with the
energy loss associated with a decay event.
This kind of issue has been encountered before in association with loss in the case of
quantum optical soliton propagation in a fiber.\cite{Fini}
Photons experience an attractive interaction through the fiber nonlinearity, 
which counteracts dispersion associated with variations in the index of refraction,
to produce correlated many-photon states (optical solitons).
Photons are lost when absorbed or scattered by the fiber.
When a photon is absorbed, the wavefunction of the remaining system is impacted,
leading to increases in the center of mass position and momentum.
This leads to quantum noise associated with the optical soliton, and is termed the
Gordon-Haus effect.
An analysis of this effect in the Schr\"odinger picture was recently presented by Fini et al.\cite{Fini}

In this analysis, photon loss is included through matrix elements between
initial and final correlated many-photon states which differ in photon number and energy.
Underlying this approach is a recognition that most of the correlations present before the
decay process will remain after the decay.
The situation is similar in the present case.
Prior to a decay, we have a coupled system with correlations between the different 
basis states.
After the decay has occurred, the coupled system remains coupled, and whatever correlations
were present initially should for the most part still be present.
Consequently, the probability of a particular loss event can be determined in terms of 
initial and final states that include these correlations.

  This argument implies that we should calculate the rate of dissipation at a specific
energy exchange with a Golden Rule rate formula expressed in terms of eigenfunctions of the coupled system

\begin{equation}
\Gamma(E_b-E_a)
~=~
{2 \pi \over \hbar}
\sum_j
|
\langle \Phi(E_b) \psi(\epsilon_b)
|\hat{U}|
\Phi(E_a) \psi(\epsilon_a)
\rangle_j |^2
\rho(E_f)
\end{equation}

\noindent
Here the initial nuclear state at site $j$ is $\psi(\epsilon_a)$ with energy $\epsilon_a$; 
the final nuclear state after disintegration is $\psi(\epsilon_b)$ with energy $\epsilon_b$.
The initial state of the coupled oscillator and two-level systems is $\Phi(E_a)$ with energy $E_a$;
the final state of the coupled system is $\Phi(E_b)$ with energy $E_b$.
For loss to occur, the coupled system lowers its energy from $E_a$ to $E_b$, and the nucleus
accepts the energy difference, so that

\begin{equation}
\epsilon_b ~=~ \epsilon_a + (E_b-E_a)
\end{equation}

\noindent
The dissipation comes about in this case when two eigenfunctions of the coupled system $\Phi(E_a)$ and $\Phi(E_b)$ 
that have a large energy difference are connected by a low-order phonon operator.

\subsection*{Loss avoidance and approximate wavefunctions}

We discussed above the issue of loss avoidance in the presence of strong loss in the case
of two generic levels, and we found that the probability amplitude avoided regions of high
loss in the case of the low-loss solution.
We expect the coupled system composed of a highly excited oscillator and two-level systems
to behave similarly.
We are interested here in the question of how the coupled system accomplishes this with
loss models similar to those considered in Appendix A.
  From the discussion above, loss can be minimized when the magnitude of the matrix
element involved is minimized.
In the event that we use a disintegration model based on dipole coupling (as proposed in 
Appendix A), then this matrix element is proportional to

\begin{equation}
\bigg \langle \Phi(E_b)\bigg |
\langle  \psi(\epsilon_b)
|\hat{U}|
 \psi(\epsilon_a)
\rangle 
\bigg | \Phi(E_a) \bigg \rangle
~\sim~
\bigg \langle \Phi(E_b)\bigg |
(\hat{a}+\hat{a}^\dagger)
\bigg | \Phi(E_a) \bigg \rangle
\end{equation}

\noindent
If the loss increases rapidly with increasing energy (due to enhanced tunneling,
opening of new channels, or statistical factors), then the coupled system can
minimize loss by reducing the magnitude of such matrix elements when
the energy difference $E_b-E_a$ is large.

\epsfxsize = 3.00in
\epsfysize = 2.25in
\begin{figure} [t]
\begin{center}
\mbox{\epsfbox{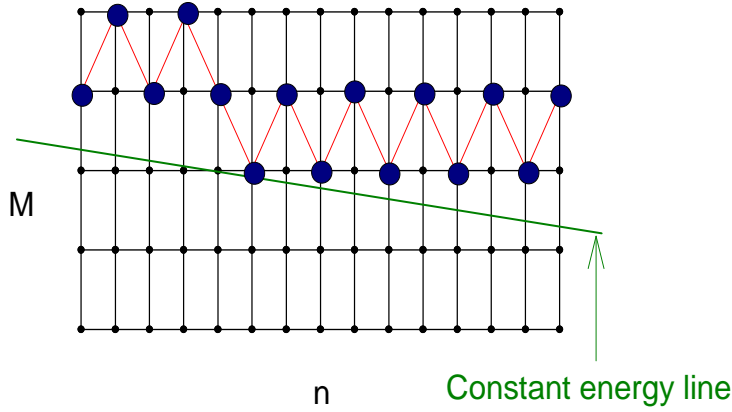}}
\caption{Schematic of wavefunction and constant energy line in the case of one set of two-level
systems coupled to an oscillator when loss increases as a strong function of the
available energy. 
This situation corresponds to intermediate coupling, in which the wavefunction includes
many points in $(n,M)$ space, which lie above the constant energy line.
This constant energy line here again corresponds to constant $M \Delta E + n \hbar \omega_0$
plus a self-energy term which we assume is small.
Loss is minimized through localization of the wavefunction in $M$, since the coupling between
similar eigenfunctions that are offset by two units in $M$ is minimized.}
\label{constE2}
\end{center}
\end{figure}

A simple approximation can be developed by working with wavefunctions for which
most of these matrix elements are identically zero for an energy exchange of $2 \Delta E$ or greater.
Such a wavefunction is illustrated in Figure \ref{constE2}.
This wavefunction lies above the constant energy line, and is restricted to having
only a single nonzero point for each $M$.
Such wavefunctions will exhibit no loss for most transitions with $E_b-E_a \ge 2 \Delta E$, but will
have loss at lower energy separation.
For example, a displacement down by one unit in $M$, and over by one unit in $n$, 
produces a lower energy wavefunction for which a finite loss matrix element 
occurs.
There is no possibility of a wavefunction construction for which finite coupling 
is maintained and this kind of loss can be avoided completely.
However, the magnitude of the loss for one unit separation in $M$ will depend on
the relative phases of the wavefunction in $n$, so that in a subset of the
possible solutions this loss will be minimized.

   In the case of a single oscillator coupled to two sets of two-level systems,
a similar construction is possible in which the wavefunctions lie above and
close to the constant energy surface.
Such solutions again minimize the loss in the presence of dissipation that increases
strongly with energy.
In the numerical calculations of the coupled system with a lossy oscillator, we have
made use of these kinds of approximate wavefunctions.


\newpage
\section*{Appendix C: Energy exchange with the oscillator}

In the models under discussion in the main text, the two-level systems on the 
receiver side are strongly coupled to the oscillator.  This coupling produces 
a self-energy shift as well as an energy exchange effect with the lattice.  In 
this appendix we develop an approximate solution for the coupled lossy oscillator and 
two-level systems on the receiver side in order to study energy exchange between
the two-level systems and the oscillator.
  In the presence of loss, the receiver-side sector Hamiltonian is

{\small

\begin{equation}
\hat{H}_2
~=~
\Delta E_2  {\hat{S}_z^{(2)} \over \hbar}
+
\hbar \omega_0 \hat{a}^\dagger \hat{a}
-
{i \hbar \over 2}\hat{\Gamma}(E)
+
V_2 {2 \hat{S}_x^{(2)} \over \hbar}(\hat{a}+\hat{a}^\dagger)
\end{equation}

}

\noindent
where as before we assume that the oscillator energy $\hbar \omega_0$ is small compared to the
two-level system energy $\Delta E_2$.  We are interested in this problem when $n$ is large,
when the number of two-level systems is large ($S_2 \gg 1$), and when there is significant
excitation of the two-level systems.

  We seek approximate solutions to the associated time-independent Schr\"odinger
equation to gain some understanding of the system dynamics.  
We adopt a loss model of the kind discussed in Appendix A, with the loss increasing
rapidly with increased energy exchange.
We adopt approximate solutions that are localized in $M_2$ as discussed in Appendix B.
The loss in the vicinity of $\Delta E_2$ in what follows is taken to be weak
(this assumes that the Dicke enhancement factors associated with coherence are
large).
Although we have studied this kind of problem using both analytic and numerical
approaches, a presentation of the results would make this manuscript excessively
long.
We instead will focus here on an approximate local solution, which appears to
capture many important features observed in numerical calculations and analytical
work.

Wavefunction solutions in $(n,M_2)$ space lie above and close to a constant energy line
as illustrated in Figure \ref{constE2} in the model under consideration.
In this case, we can develop an approximate local solution of the form

\begin{equation}
\Psi(M_2,n)
~=~
e^{in \theta}
\times
\left \lbrace
\begin{array} {cc}
-\delta_{M_2-M_0} & {\rm even} ~ n \cr
\delta_{M_2-M_0-1} & {\rm odd} ~ n \cr
\end{array}
\right .
\end{equation}

\noindent
where $M_0$ is the smallest local value of $M_2$ that lies above the
constant energy line (we have used even and odd $n$ here to implement
solutions that alternate in $n$; similar results are obtained with
even and odd reversed here).  
This solution is to be considered to be local in the sense that we will adopt
it only in the vicinity of $(n_0, M_0)$. 
Associated with this local solution is an energy eigenvalue estimate 

\begin{equation}
E_2(\theta) 
~\approx~
\Delta E_2 M_0
~+~
\hbar \omega_0 n_0 
~-~
2 V_2 \sqrt{n_0} \cos(\theta) \sqrt{S_2^2 - M_0^2}
\end{equation}

\noindent
For this approximate local solution to be good, we need to be locally in the
strong coupling limit

\begin{equation}
2 V_2 \sqrt{n_0} \cos(\theta) \sqrt{S_2^2 - M_0^2}
~\gg~
\Delta E_2
\end{equation}

\noindent
In the event that dissipation comes about due to dipole coupling as discussed
in Appendix A, then the dissipation rate in this model will be proportional
to 

\begin{equation}
\Gamma(\theta) ~\sim~ 1 - \cos (\theta)
\end{equation}

\noindent
The lowest dissipation then appears when $\theta=0$.

This local solution has given us an estimate of the self-energy associated
with rapid excitation transfers in the receiver system in the strong coupling limit.
From the dependence of the self-energy on $\theta$, we can determine roughly how
fast energy can be exchanged between the oscillator and two-level systems.
For example, the associated group velocity for energy exchange with the oscillator in
this model is 

\begin{equation}
{d n \over dt}
~\sim~
{1 \over \hbar} {d E_2 \over d \theta}
~=~
{2 V_2 \sqrt{n_0} \sin (\theta) \sqrt{S_2^2 - M_0^2} \over \hbar}
\end{equation}

\noindent
This rate can be very fast.  
The coupled oscillator and two-level systems on the receiver side appear to be 
capable of efficient coherent energy exchange under conditions where the two quantum 
systems have incommensurate energy quanta.
If the loss is minimized at the value of $\theta$ in which the velocity is zero,
as occurs in the simple loss model we considered here, then one would expect
to see dissipation in association with the coherent energy exchange.


\newpage
\section*{Appendix D: Excitation transfer}

We are interested in this appendix in examining excitation transfer in the strong coupling
regime, including the effects of loss.
We recall that in the model under discussion, two sets of two-level systems are coupled
to a highly excited oscillator.
When no loss is present the excitation transfer effect is weak, and it requires that
the energies be closely matched (as discussed in the main text).
When the model is augmented with loss, the situation changes.
The excitation transfer effect is strongly enhanced in the weak coupling regime and
intermediate coupling regime as discussed in the main text.
We expect that the effect is similarly enhanced in the strong coupling regime, which
motivates us to explore a simple local model for excitation transfer here.

  We have also seen that the introduction of loss has a strong effect on energy 
exchange between the two-level systems and oscillator on the receiver side, 
significantly increasing the strength of the effect.
We noted in the case of our numerical results in the intermediate coupling regime 
that there occured a weak energy exchange effect in association with excitation
transfer, such that the oscillator was able to make up the difference between the
dressed two-level energies. 
Here we are interested in the possibility of energy exchange with the oscillator in association
with excitation transfer between the two sets of two level systems under conditions where
a mismatch occurs.

  To investigate these issues, we begin with the model Hamiltonian

{\small

\begin{equation}
\hat{H}
~=~
\Delta E_1  {\hat{S}_z^{(1)} \over \hbar}
+
\Delta E_2  {\hat{S}_z^{(2)} \over \hbar}
+
\hbar \omega_0 \hat{a}^\dagger \hat{a}
-
{i \hbar \over 2}\hat{\Gamma}(E)
+
V_1 e^{-G} {2 \hat{S}_x^{(1)} \over \hbar}(\hat{a}+\hat{a}^\dagger)
+
V_2 {2 \hat{S}_x^{(2)} \over \hbar}(\hat{a}+\hat{a}^\dagger)
\end{equation}

}

\noindent
We are interested in the development of an approximate local solution 
in $(n,M_1,M_2)$ space.  
We have discussed loss models in Appendix A, and we introduced an approximate
wavefunction that minimizes loss when the loss increases rapidly with
increasing energy exchange in Appendix B.
When the coupling is weak, this wavefunction consists of a set of points in $(n,M_1,M_2)$
space that lie above a constant energy surface

\begin{equation}
E ~=~ \Delta E_1 M_1 + \Delta E_2 M_2 + \hbar \omega_0 n + \Sigma(E)
\end{equation}

\noindent
When the coupling is strong the self-energy $\Sigma(E)$ may be large, so that the
constant energy surface may move away from its weak-coupling location.
Nevertheless, a wavefunction in the strong coupling limit would be expected to
lie above a surface, and localize close to this surface in order to minimize
loss.

  We adopt a local solution of the form

\begin{equation}
\Psi(M_1,M_2,n)
~=~
e^{i n \theta}
e^{i(M_2-M_1)\phi}
\left \lbrace
\begin{array} {ll}
- \cos (\alpha) \delta_{M_1+M_2-M_0}   & {\rm even}~ M_0, {\rm even}~n \cr
\sin (\alpha) \delta_{M_1+M_2-M_0-1}  & {\rm odd}~  M_0+1, {\rm odd}~ n \cr
\end{array}
\right .
\end{equation}

\noindent
This solution is constructed to lie locally above a constant energy surface (when $\Delta E_1$ is
roughly matched to $\Delta E_2$), consistent with the loss-minimizing solution discussed in Appendix B.
As before, we consider this solution to be local in the sense that we use it only in the vicinity
of a point in $(n,M_1,M_2)$ space.
An approximate energy eigenvalue can be developed from the solution of the determinantal
equation

\begin{equation}
\det
\left (
\begin{array} {cc}
E_0 - E & v_1(\theta,\phi)+v_2(\theta,\phi)   \cr
v_1^*(\theta,\phi)+v_2^*(\theta,\phi) & E_0 + \Delta E - E \cr
\end{array}
\right )
~=~
0
\end{equation}

\noindent
We have defined $E_0$ according to

\begin{equation}
E_0 ~=~ (\Delta E_1 M_1 + \Delta E_2 M_2 + \hbar \omega_0 n)_0
\end{equation}

\noindent
We assume that the two-level systems are resonant:

\begin{equation}
\Delta E ~=~ \Delta E_1 ~=~ \Delta E_2
\end{equation}

\noindent
The interactions $v_1$ and $v_2$ are given by

\begin{equation}
v_1(\theta,\phi)
=
2 \sqrt{n_0} \cos(\theta) 
 V_1 e^{-G} e^{-i \phi} \sqrt{S_1^2-(M_1)_0^2} 
\ \ \ \ \ \ \ 
v_2(\theta,\phi)
=
2 \sqrt{n_0} \cos(\theta) 
 V_2 e^{i \phi} \sqrt{S_2^2 - (M_2)_0^2}  
\end{equation}

\noindent
The general formula for $E(\theta,\phi)$ that results is complicated, and not particularly
interesting for our present discussion.  More interesting is the 
formula obtained in the limit that the coupling on the receiver-side
is strong.  In this limit we may write

{\small

\begin{equation}
(E_0-E)(E_0+\Delta E-E)
-
|v_2|^2 
\left( 
1 + {v_1 \over v_2} + {v_1^* \over v_2^*}
\right )
= 0
\end{equation}

}

\noindent
There are two band solutions since this equation is quadratic.  We take
the lower energy band since it has a negative self-energy, in 
keeping with the weak coupling limit

$$
E(\theta,\phi)
~\approx~
E_0 + {\Delta E \over 2}
-
|v_2| \sqrt{1+{v_1 \over v_2} +{v_1^* \over v_2^*} }
$$

\noindent
We can then use a Taylor series approximation to obtain

\begin{equation}
E(\theta,\phi)
~\approx~
E_0 + {\Delta E \over 2}
-
2 \sqrt{n_0} \cos(\theta)
\left [
V_2  \sqrt{S_2^2 - (M_2)_0^2}
 +
V_1 e^{-G} \cos(2 \phi) \sqrt{S_1^2-(M_1)_0^2}
\right ]
\end{equation}

\noindent
We see in this formula a receiver-side self-energy term which is the
same as what we found in Appendix C.  In addition, we obtain a new
term from which we are able to determine the group velocity associated
with excitation transfer.

There are two features of this result that are of interest to us.
In the event that the two-level systems on the donor and receiver
sides are matched, then the associated rate for coherent excitation
transfer between the two sides is

\begin{equation}
{d (M_2-M_1) \over dt}
~\sim~
{1 \over \hbar} {d E \over d \phi}
~=~
{4 V_1 e^{-G} \sqrt{n_0} \cos(\theta) \sin (2 \phi) \sqrt{S_1^2 - (M_1)_0^2} \over \hbar}
\end{equation}

\noindent
This result is of interest since it is sufficiently large to be relevant
to experiments that show an excess heat effect (excitation transfer from D$_2$ in
this model is assumed to be followed by either coherent or incoherent
energy exchange with the lattice).
The second reason that this result is of interest to us is that it suggests 
that we might expect roughly the same rate for excitation transfer if the donor and receiver
levels are not precisely matched, since solutions may be oscillatory in both
$\phi$ and $\theta$. 
We recall that in the absence of loss we required a precise resonance since the
coupling was weak. 
The inclusion of loss results in much stronger coupling, and in the limit that the
coupling is strong, then the system appears to be able to evolve with its maximum
rate even when a precise resonance is absent.
This is a dramatic result, and one which we investigate further in Section VI.


\newpage

\section*{Acknowledgments}
Support for P. L. Hagelstein was provided by Spindletop Corporation.
Support for I. Chaudhary was provided by the Kimmel Fund, the Bose Foundation,
the Griswold Gift Program and by a DARPA subcontract.


\end{document}